\documentclass [12pt]{article}

\usepackage{bm}
\usepackage{float}
\restylefloat{table}
\restylefloat{figure}

\usepackage{upquote}
\usepackage{graphicx,amssymb,amsbsy,amsfonts,amssymb,amsmath}
\usepackage[usenames, dvipsnames]{xcolor}
\usepackage{calc}
\usepackage{url}

\allowdisplaybreaks
\usepackage{array}
\usepackage{bbm}
\usepackage{slashed}

\numberwithin{equation}{section}

\newif\ifdraft
\drafttrue
\newif\ifpreprint
\preprinttrue

\restylefloat{figure}
\allowdisplaybreaks

\renewcommand{\theequation}{\thesection.\arabic{equation}}

\newcommand{\req}[1]{(\ref{#1})}
\def\vev#1{\langle #1 \rangle}

\def\fc#1#2{\frac{#1}{#2}}
\def\h{\frac{1}{2}}
\newcommand{\nwc}{\newcommand}
\nwc{\ba}  {\begin{array}}
\nwc{\ea}  {\end{array}}
\nwc{\bdm} {\begin{displaymath}}
\nwc{\edm} {\end{displaymath}}

\nwc{\bea} {\begin{equation}\ba{lcl}}
\nwc{\eea} {\ea\end{equation}}

\nwc{\be} {\begin{equation}}
\nwc{\ee} {\end{equation}}

\nwc{\bda} {\bdm\ba{lcl}}
\nwc{\eda} {\ea\edm}

\nwc{\bc}  {\begin{center}}
\nwc{\ec}  {\end{center}}

\nwc{\ds}  {\displaystyle}

\nwc{\nn} {\nonumber}
\nwc{\nnn} {\nonumber \vspace{.2cm} \\ }
\nwc{\ra}{\rightarrow}
\nwc{\lra}{\longrightarrow}
\def\lf{\left}\def\ri{\right}

\nwc{\p} {\partial}
\nwc{\Tr}{{\rm Tr}}
\def\IR{{\bf R}}

\def\ap{\alpha'}

\def\Mc{{\cal M}}

\def\ov{\overline}

\def\eps{\epsilon}

\newcommand{\floor}[1]{\lfloor #1 \rfloor}

\begin{document}

\title{\textbf{Black Hole Formation and Classicalization 
in  Ultra-Planckian $\bm{2\rightarrow N}$ Scattering}  \\ }
\author{\large G. Dvali$^{\textrm{a,b,c}}$,  C. Gomez$^{\textrm{a,d}}$, 
R.S. Isermann$^{\textrm{a}}$, D. L\"ust$^{\textrm{a,b}},$ and  
S. Stieberger$^{\textrm{b}}$\\[2cm]}
\date{}
\maketitle
\vskip-2.5cm
\centerline{\it $^{\textrm{a}}$ Arnold--Sommerfeld--Center for Theoretical Physics,}
\centerline{\it Ludwig--Maximilians--Universit\"at, 80333 M\"unchen, Germany}
\medskip
\centerline{\it $^{\textrm{b}}$ Max--Planck--Institut f\"ur Physik,
Werner--Heisenberg--Institut,}
\centerline{\it 80805 M\"unchen, Germany}
\medskip
\centerline{\it $^{\textrm{c}}$ Center for Cosmology and Particle Physics,
Department of Physics, New York University}
\centerline{\it 4 Washington Place, New York, NY 10003, USA}
\medskip
\centerline{\it $^{\textrm{d}}$ Instituto de F\'{\i}sica Te\'orica UAM-CSIC, C-XVI,
Universidad Aut\'onoma de Madrid,}
\centerline{\it Cantoblanco, 28049 Madrid, Spain}

\begin{abstract}
We establish a connection between the ultra--Planckian scattering amplitudes  
in field and string theory and unitarization by black hole formation in these scattering processes.  
Using as a guideline an explicit microscopic theory in which the black hole represents a bound-state of many soft gravitons at the quantum critical point,  we were able to identify and compute a set of perturbative amplitudes relevant for black hole formation. 
 These are the tree--level $N$ graviton scattering $S$--matrix elements in a kinematical regime (called classicalization limit) where the two incoming ultra-Planckian gravitons produce a large number $N$ of soft gravitons.  We compute these amplitudes by using  the
 Kawai--Lewellen--Tye relations, as well as scattering equations and  string theory techniques. 
  We discover that this limit reveals the key  features of the microscopic  corpuscular black hole $N$--portrait. In particular,  the perturbative suppression factor of a $N$-graviton final state, derived from the amplitude, matches the non-perturbative black hole entropy when $N$ reaches the quantum criticality 
value, whereas final states with different value of $N$ are either suppressed or excluded by non-perturbative 
corpuscular physics.  Thus we identify the microscopic reason behind the  black hole dominance over other final states including non-black hole classical object. In the parameterization of the classicalization limit the scattering equations can be solved exactly allowing us to obtain closed expressions for the high--energy limit of the open and  closed superstring tree--level scattering amplitudes for  a generic number $N$ of external legs.
We demonstrate matching and complementarity between the string theory and field theory in different 
large-$s$ and large-$N$ regimes. 

\end{abstract}

\begin{flushright}
{\small  MPP--2014--320}\\
{\small LMU--ASC 52/14}
\end{flushright}


\thispagestyle{empty}

\newpage
\setcounter{tocdepth}{2}
\tableofcontents
\break

\section{Introduction and summary} 

    The formulation of a microscopic picture of black hole production
    in high--energy particle scattering 
 is crucial for understanding the nature of quantum gravity at ultra-Planckian energies. 
 In particular, this issue is central to the idea that gravity is UV-complete in a non-Wilsonian 
sense \cite{Self}, based on the concept of classicalization \cite{Dvali:2010jz}.  

The standard (Wilsonian) approach to UV-completion implies that 
 interactions at higher and higher energies are regulated by integrating-in weakly-coupled degrees of freedom of shorter and shorter wave-lengths.  
  When applied to gravity,  the Wilsonian picture would imply that  at energies exceeding the Planck mass, $\sqrt{s} \, \gg \, M_P$,  the UV-completion must be achieved by new quantum  degrees of freedom of wavelength 
  much shorter than the Planck length,  $R \, \sim {1 \over \sqrt{s}} \, \ll \, L_P$. 
  In the classicalization approach, instead of introducing new hard quanta, the UV-completion is accomplished by means of {\it collective} states composed of  a
 large number $N \sim s/M_P^2$ of soft gravitons  of wavelength $R \sim \sqrt{N} L_P$ \cite{Dvali:2011th}
 that, in the mean-field approximation, recover the semi-classical behavior of macroscopic black holes \cite{Dvali:2011aa}.  To put it shortly, classicalization replaces the hard  quanta  by a multiplicity of soft ones, 
 which in mean-field (large $N$) approximation acquire  some properties of classical objects.   
 
     In the conventional semi-classical approach, the current understanding of black hole production is rather unsettling.  On one hand, it is widely accepted  
     that scattering of very highly energetic particles results into a black hole formation.   This acceptance is based on the following argument:     
     according to classical gravity any source of center of mass energy $\sqrt{s}$  
 when localized within its gravitational (Schwarzschild) radius  $R = \sqrt{s}L_P^2$ must form a black hole. 
   This argument is insensitive to the precise nature of the source and in particular should be applicable 
   to elementary particle sources.  Thus,  it is reasonable to expect that, for example, a two-particle scattering with center of mass energy 
of the order of the solar mass for an impact parameter less than $3$km, should result into the formation 
of a solar mass black hole.     

 On the other hand,  we have to admit that this way of thinking challenges the view about black holes as {\it classical}  macroscopic objects, since production of usual macroscopic objects in two-particle collisions is expected to be exponentially-suppressed.  For example, in the above thought experiment of two-particle collision at solar mass energy it is exponentially-unlikely for a sun-like object to be produced in the final state instead of a black hole. 
 
  What makes black holes so different from ordinary classical objects from the point of view of their microscopic structure?
  
%
  Of course, one can certainly say that what makes black holes very special is their Bekenstein--Hawking entropy.  
 However, without a microscopic explanation of entropy creation in two-particle collision, this invocation of the entropy is only making 
the  puzzle more complicated.  Indeed, it is totally mysterious how an initial two-particle state with zero entropy gains such an enormous entropy 
 in the process of the collision. 
 
      The above questions are impossible to answer without having a microscopic theory of the black hole  
 and the corresponding microscopic mechanism of black hole formation in particle scattering processes.   
   This is why  the above questions have not been settled although the study of black hole formation in particle collisions at ultra-Planckian energies has been pioneered long ago \cite{ultra1,  Amati:1987wq, Gross:1987kza} and since then has even been taken as far as predicting production of micro black holes at LHC \cite{LHC}. 
   The reason is the lack of a quantum corpuscular picture of black holes which subsequently 
   makes it impossible to figure out how  the quantum gravity amplitude translates into the formation of a black hole 
   final state.

    The present paper is an attempt to establish the missing link between quantum gravity amplitudes and a corpuscular picture of black holes.
  In particular we  will provide the link between the
  corpuscular black hole portrait \cite{Dvali:2011aa} on the one hand and the classicalization idea for gravitational scattering amplitudes\footnote{There is another
  attempt for a synthesis   \cite{Kuhnel:2014xga} by sewing together two $2 \rightarrow N$ graviton amplitudes into a ladder loop diagram and coherently summing over different $N$ in an eikonal 
  region.}.
    By employing the 
  corpuscular black hole picture together with the expressions of graviton scattering amplitudes both in field and string theory we shall  uncover some key elements underlying the microscopic  origin of black hole formation.

    More concretely: 
 
 \begin{itemize} 
 
  \item Guided by non-perturbative input from the corpuscular black hole $N$-portrait, 
  we identify the black hole formation regime as the regime 
  of multi-particle creation,  in form of  $2 \rightarrow N$ graviton scattering amplitudes,  with number of soft gravitons in the final state being given by the number of black hole constituents, as suggested by
 classicalization. 
 
 \item Next, by using powerful field and string-theoretic techniques, in particular 
 scattering equations \cite{Cachazo} and Kawai-Lewellen-Tye (KLT)  relations \cite{Kawai:1985xq}, we estimate the perturbative part of these $N$-graviton amplitudes. 
 
    \item  Finally, using the microscopic corpuscular picture of black holes as $N$-graviton self-bound 
    states at a quantum critical point, we provide the missing non-perturbative information that enables us to translate the 
$N$-graviton production processes into the black hole formation, both in field and string theory
    scatterings. 
    
    \item We provide a cross-check of perturbative $N$-graviton amplitudes by applying them 
 to the production of non-black hole type classical configurations described by multi-particle coherent states for which semi-classical estimates must also be valid.   We then match the two results
 and  observe that the exponential suppression expected in the semi-classical theory is indeed reproduced by 
 the perturbative $2\rightarrow N$ gravity amplitudes. 
 Thus, this matching besides of providing 
 an independent information about the multi-graviton amplitudes,  also confirms that the microscopic 
  origin of the black hole dominance, relative to other possible multi-particle final states of the same energy, lies in the quantum criticality
  of the black hole constituents, which is absent for other classical objects.    
  
  \item One of the outcomes of our analysis is to show the very different large-$N$ behavior  of multi-particle amplitudes in gravity in comparison with non-derivatively coupled scalar theories.

    \end{itemize} 
    
   The above framework supplies a correct physical picture that among other things explains why the black hole production is the dominant   process  while the production of other macroscopic multi-particle states is exponentially-suppressed.   
   The perturbative kinematics that we shall identify  has just the right suppression to be compensated by the
  degeneracy of states at the quantum critical point. In other words, in this multi particle production kinematics, the amplitude itself anticipates what would be the right value for the entropy.  

We also observe a nice interplay between the field and string theory amplitudes.  
 In particular, we observe that the  string and field theory amplitudes agree  whenever the size of the produced black hole is larger than the string length, or equivalently, when the Reggeization of the amplitude does not take place.       
 
  Before moving into the technical part of the paper, to be covered in the following sections, we shall summarize 
 the basic results and their physical meaning.  
 In order to do it we shall briefly review the non-perturbative input coming from the corpuscular  black hole portrait, 
  which being a microscopic quantum  theory, provides a crucial missing link between the perturbative $N$-graviton production  amplitudes and the unitarization of the theory by black hole formation.

 \subsection{Non--perturbative input from a microscopic portrait} 
 
  In order to make the connection explicit let us summarize some non-perturbative input 
 coming from the black hole corpuscular quantum portrait \cite{Dvali:2011aa, GiaCesarQC} (for other aspects of this proposal see 
 \cite{NportraitO} and some similarities with this proposal can be found in \cite{NportraitOa}).   
    This  portrait is based on the idea that the black hole is a composite entity. 
   Its corpuscular constituents are gravitons  
  with the characteristic de Broglie wavelength given by the classical size of the black hole,  $R$.  That is, 
  the internal (and near-horizon) physics of black holes is fully determined by the quantum interaction 
  of gravitons of wave-length $R$.  We shall be interested in the regimes in which the black hole is much heavier than the Planck mass $M_{BH}  \gg \, M_P$, or equivalently,  $R \, \gg \,  L_P$.  
   
   The two crucial properties  are:  
   
   \begin{itemize} 
   
   \item  For macroscopic black holes,  the quantum gravitational coupling $\alpha$ among the individual corpuscles,  
   \begin{equation}
   \alpha \, \equiv {L_P^2 \over R^2}
   \end{equation}  
   is extremely weak.
   
   \item The number $N$ of constituents of wavelength $R$ is:  
   \begin{equation}
   N \, =\,  M_{BH}^2/M_P^2\, .
   \end{equation}

    \end{itemize} 
  
   Thus,  quantum-mechanically  a black hole represents a self-bound state of  soft gravitons, with 
    a very special interplay between the quantum coupling and the number of constituents, $\alpha N \, = \, 1$.   Or equivalently, the black hole  is a state in which the wave-lengths of gravitons 
    satisfy,   $R \, = \, \sqrt{N} L_P$.  
    This property implies that the physics of black holes is similar to that of a graviton 
   Bose-Einstein  condensate  at a {\it quantum critical point} \cite{GiaCesarQC}.   This  critical point separates the following two phases.   For $\alpha N \ll 1$,  the system is in the phase in which collective graviton-graviton attraction is not enough to form a self-bound-state and the graviton Bose-gas is essentially free.  For $\alpha N =1$ the bound-state is formed.    
     
   At this critical point,  order
   $N$ collective Bogoliubov modes become gapless leading to an exponential degeneracy 
   of states, of order $e^{cN}$, where $c$ is some positive constant.  This exponential degeneracy of states is quickly lifted when we deform the system and move away from the critical point $\alpha N \neq 1$.  
   While for the generic attractive Bose-gas, moving away from the critical point is possible in both directions,
  ($\alpha N < 1$  or  $\alpha N > 1$), for gravitons this is not the case.
  The gravitons cannot form 
  a sensible state with $\alpha N \, \gg 1$.  \footnote{In more than one space dimensions the attractive Bose-gas
  in the over-critical phase undergoes a quantum collapse. However, for gravity this is impossible since 
  gravitons of a given energy cannot form a configuration smaller than a black hole.  As result the graviton bound-state is "stuck" at the critical point slowly loosing the constituent gravitons due to the quantum depletion, reducing $N$, but maintaining quantum criticality for each $N$.  This is how the  corpuscular picture 
 accounts for the Hawking radiation.  For the purpose of the present paper, we shall ignore the 
 further evolution of black holes after their formation in the scattering process. } 
  Thus, the viability and the nature of the deformed state depends 
 in which direction we move from the critical point.

    For  $\alpha N \, < 1$, the system  of $N$ gravitons  is essentially  free. The Bogoliubov 
    frequencies are positive and Bogoliubov levels are separated by a large energy gap ($\sim  1/R$) 
  from the lowest level obtained in the free-graviton approximation.
  Due to this, the non-perturbative collective quantum effects can be ignored  and the system can be well-approximated by  an 
   asymptotic $N$-particle eigenstate of the $S$-matrix, with no additional non-perturbative information required. 
   Hence,  non-perturbative  physics gives no additional essential input for states with $\alpha N  \, \ll \, 1$, 
   and the perturbative approximation can be trusted. In particular,  the perturbative amplitudes can be directly applied to the formation of final states with  $\alpha N  \, \ll \, 1$.

   However,  for the states with $\alpha N \, > \, 1$ the situation  is very different.    The Bogoliubov frequencies of the $N$-graviton state  in this regime are complex, with the Liapunov exponent being much larger than the inverse size of the system.  This indicates that such state cannot 
 be treated as a viable asymptotic state of the $S$-matrix, even approximately.    
 This is remarkable, 
   since translated into the language of $N$-graviton perturbative amplitudes,  this regime would include the region that violates perturbative unitarity.  Thus, non-perturbative  $N$-graviton physics provides a selection rule that cuts-out  from the Hilbert space those would-be $N$-particle final states 
that perturbatively violate unitarity.  Non-perturbative corpuscular physics is telling us that 
such states are not part of the physical Hilbert space. 

The situation is schematically depicted on Fig. 1. 
The critical point $\alpha N =1$ corresponds to the point of black hole formation. At this point the levels become nearly gapless (up to $1/N$ resolution), and 
there is a maximal degeneracy of states. For estimating the production rate of $N$-graviton state 
at the critical point, the perturbative amplitudes must be supplemented by a non-perturbative factor  $\sim e^{N}$
counting the degeneracy of states, which at the same time represents the black hole entropy factor.
The region to the left corresponds to a nearly-free graviton gas. 
These states are close to asymptotic $S$-matrix states of $N$ free gravitons and their creation in two graviton collision can be estimated via perturbative matrix elements, with non-perturbative corrections being small. The region to the right is excluded by non-perturbative physics.

   The above outline summarizes the non-perturbative information that provides the missing link between the perturbative amplitude 
  and unitarization by classicalization via black hole formation.   
  
\begin{figure}[H]
\centering
\includegraphics[scale=0.45]{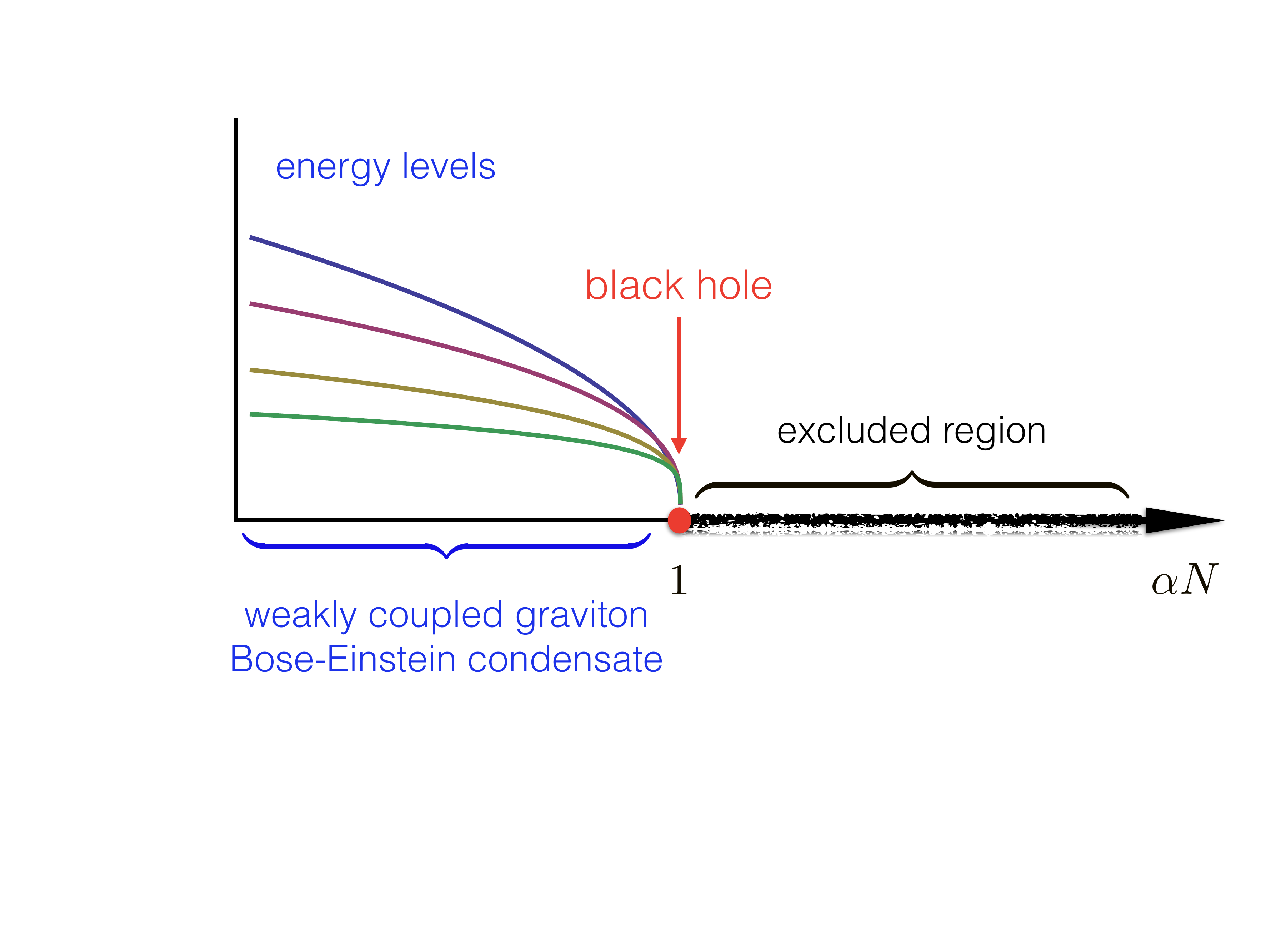}
\vskip-3cm
\caption{Bose--Einstein levels and black hole formation.}
\end{figure}

    \subsection[$N$-graviton amplitudes and black hole formation]{$\bm{N}$-graviton amplitudes and black hole formation}
  
   Based on the previous discussion the following picture emerges. In order to estimate the production rate of a $N$-graviton 
  state  we need to supplement the perturbative scattering amplitude, that views a given $N$-particle state 
  as an asymptotic state of free gravitons, by the non-perturbative information about the viability and quantum degeneracy of this state. 
  This information either will further enhance the rate or will diminish it depending where the given state is in the
  $\alpha N$ plot. 
   
   The perturbative amplitudes relevant for describing the production of a black hole of mass $M_{BH} \, = \, \sqrt{s}$ are the perturbative amplitudes at center of mass energy  $\sqrt{s}$ in which $N$-gravitons of 
 momenta $p \sim  (\sqrt{s} L_P^2)^{-1}$ are created in the final state. 
 As we shall see, 
 the transition probability of this process obtained from the corresponding  
 $S$-matrix element scales as, 
\begin{equation}
  |\langle2|S|N\rangle|^2_{pert} \sim \, \alpha^N \, N! \, =  \left ({L_P^2s \over N^2} \right )^{N} \, N! \ .    \label{pert}
  \end{equation}    
    In order to understand the physical picture, we must superimpose the non-perturbative information 
    that we have distillated from the many-body analysis of the $N$-graviton state.  Namely, the region 
   $s /M_P^2 N \, \gg \, 1$ is excluded as physically not viable due to the presence of complex Bogoliubov frequencies 
   and very large Liapunov exponent.    It is convenient to rewrite the matrix element 
   in terms of the  effective ('t Hooft-like) collective coupling, 
   \begin{equation}
   \lambda \, \equiv \, \alpha N \, = \,  s /M_P^2 N \, ,   
  \label{lambda}
  \end{equation}
  which parameterizes the strength of the collective gravitational interaction of the $N$-graviton system. 
  In this notation the matrix element becomes
  \begin{equation}
  |\langle 2 | \hat{S}| N\rangle|^2_{pert} \, \sim \,  \left ({\lambda \over N} \right )^{N} \, N!  \, .  
   \label{pert1}
  \end{equation}     
 This form makes the physics point very transparent.  
 As we just explained above, the region $\lambda \, \gg \, 1$ is excluded  by non-perturbative physics, because in this region the graviton gas is overcritical.  This includes the region in which perturbative amplitudes would violate unitarity, but multi-particle physics prevents us from going there. 
 
  The point $\lambda = 1$ is a critical point. It is allowed both perturbatively as well as non-perturbatively, but the non-perturbative 
  information is very important.   
 Notice, that for $\lambda = 1$, this amplitude has just the right scaling for being compensated by the black hole entropy factor.  In the microscopic picture 
 such a factor is indeed appearing due to the
 exponential degeneracy of states at the critical point $\lambda = 1$. 
  For values of $\lambda \ll 1$,  the system is sub-critical.  This means that not only graviton-graviton interaction is weak, $\alpha \, \ll \, 1$, but also that the collective non-perturbative effects are negligible.  Thus, the gravitons are essentially free. 
   The perturbative suppression of the amplitude cannot be compensated by the multiplicity of states, because  for $\lambda \, \ll \, 1$ the degeneracy of Bogoliubov levels is lifted and 
  there is no longer  
  an enhancement of the number of states. 
    Therefore,  using Stirling's formula and the large $N$ limit, the production rate of such multi-particle configurations, unlike black holes,  is  exponentially 
   suppressed:   
   \begin{equation}
    |\langle 2 | \hat{S}| N\rangle|^2_{pert} \,
   \sim \, e^{-N} \lambda^{N}\, .
   \end{equation}
   As already mentioned, the exponential suppression factor can be compensated by the black hole entropy.
        
      Hence, in this picture, the microscopic explanation of the black hole dominance 
 over other possible multi-particle final states, is that the latter systems are far away from 
 quantum criticality and one must pay an exponential suppression price for their production. 
  In particular,  this explains, why at a given ultra-Planckian center of mass energy $\sqrt{s}$, the production 
  rate of a non-black hole classical configuration 
  is exponentially-suppressed relative to the production rate  of a same-energy black hole. 
   The reason is that a non-black hole classical configuration of a given mass represents a coherent state of constituents that are softer and have larger occupation number than the constituents of the same mass black hole.  
 As a result such states always are at 
   the subcritical value of the collective coupling, $\lambda \, \ll \, 1$, and 
   no enhancement is available. 
    We shall discuss this point in more details towards the end of the paper, by estimating a production probability of a particular classical configuration both via the quantum $2\rightarrow N$ process  as well as semi-classically and comparing it to the  black hole production rate.

 \subsection{Various regimes} 
 
 It is instructive to summarize the various regimes of multi-particle production amplitudes, by  superimposing 
 the perturbative and non--pertubative inputs. In doing so we shall use both field theory as well as string theory data, namely,  $s,L_P, L_s$ and $N$. 
 
  Let us consider first the perturbative input.  Here we can distinguish two regimes. 
 
  \begin{itemize} 
  
  \item  The stringy regime is achieved for  
  \begin{equation}
  {sL_s^2 \over N^2} \, > \, 1\, .
  \label{stringyregime}
 \end{equation}
 This is the regime for which  the amplitudes effectively Reggeize.  
 
 \item  The field theoretic regime is achieved for, 
  \begin{equation}
  {sL_s^2 \over N^2} \, < \, 1 \,.
  \label{fieldregime}
 \end{equation}
In this case,  we can use both scattering equations and the KLT prescription and the amplitudes computed within field and string theory  agree. 

 \end{itemize} 

     On the above perturbative information we need to superimpose the non-perturbative input 
     coming from the black hole's corpuscular portrait.   This gives the following three regimes
     
   \begin{itemize} 
  
  \item  The regime of black hole formation at the critical point $\lambda = 1$,  
  \begin{equation}
  {sL_P^2 \over N} \, = \, 1\, .
  \label{criticalregime}
 \end{equation}
 In this regime the perturbative amplitudes must be supplemented by the information about the exponential multiplicity 
 of the final states.  
 
 \item   Sub-critical regime $\lambda <1$, 
  \begin{equation}
  {sL_P^2 \over N} \, < \, 1 \,.
  \label{sub-criticalregime}
 \end{equation}
 In this regime the Bogoliubov degeneracy is lifted and the perturbative amplitudes are better and better applicable as we move towards  $\lambda \, \rightarrow \, 0$. 
 \item   Over-critical regime $\lambda \, > \, 1$, 
  \begin{equation}
  {sL_P^2 \over N} \, > \, 1 \,.
  \label{overcritical}
 \end{equation}
In this regime,  the $N$-graviton  states  have very high Liapunov exponents and imaginary 
Bogoliubov frequencies and therefore are not legitimate final states. 

 \end{itemize}     

 The overlap of the perturbative and non-perturbative information is summarized 
 in the plots on Fig. 2 and 3, where various perturbative and non-perturbative regimes are plotted 
 on the $\lambda$ axis.  In order to allow the variation of $\lambda$, we vary $N$ while keeping 
 $s,L_P$ and $g_s$ fixed.  In this way, we scan all possible multi-graviton final states of the desired 
 kinematical regime for the fixed center of mass energy. 
 
The first plot on Fig. 2 describes various regions on the $\lambda$-axis from purely field-theoretic perspective
  of non-perturbative $N$-graviton physics.  
 The second plot, describes the interplay between the string-theoretic and field-theoretic domains from purely perturbative perspective.    Notice, that after translating (\ref{stringyregime}) in terms of the string 
 coupling, the transitional point between the string and field-theory regimes is 
 marked by $\lambda \, = \, Ng_s^2$. This makes a nice physical sense. Specifically, the stringy regime becomes important when the gravitational coupling between the constituents becomes weaker than the string coupling. 
  The same plot indicates the obvious point, that for any fixed value of $g_s$, and for sufficiently large $N$ 
  the field theory regime becomes a good approximation.

\begin{figure}[H]
\centering
\includegraphics[scale=0.35]{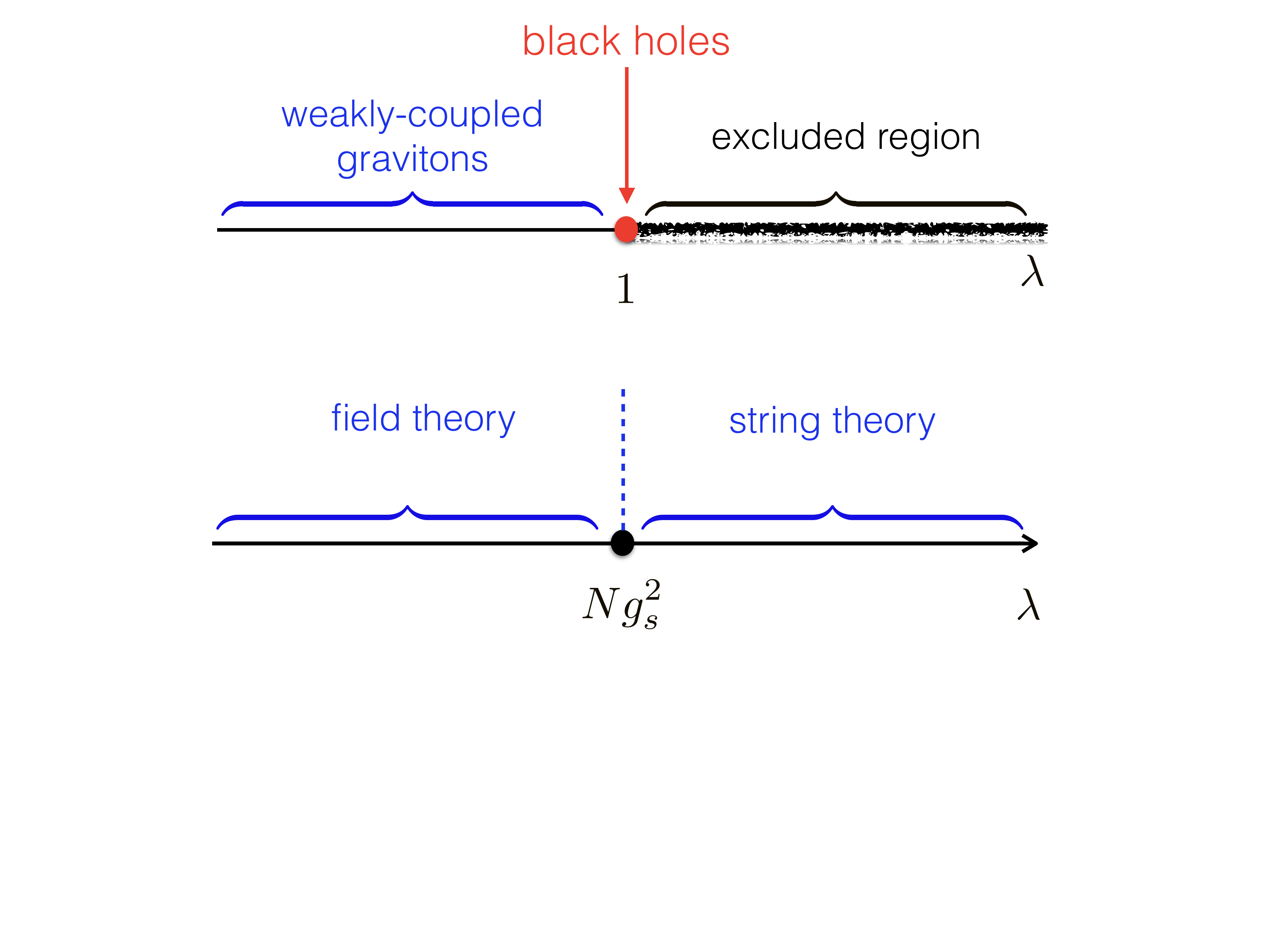}
\vskip-3cm
\caption{Graviton physics and interplay between field and string theory as variation of 
$\lambda$.}
\end{figure}

Fig. 3 described the result of super-imposing the above two perturbative and non-perturbative plots for two different cases, $Ng_s^2 > 1$ and $Ng_s^2 < 1$ respectively. In the first case, there is a region,
  $1 < \lambda < \, Ng_s^{2}$, in which on one hand unitarity is perturbatively violated in field theory and on the other hand perturbative string theory corrections are not effective for restoring it. In this domain unitarity is restored by non-perturbative collective $N$ graviton physics described above, which excludes 
  this region as unphysical.

\begin{figure}[H]
\centering
\includegraphics[scale=0.35]{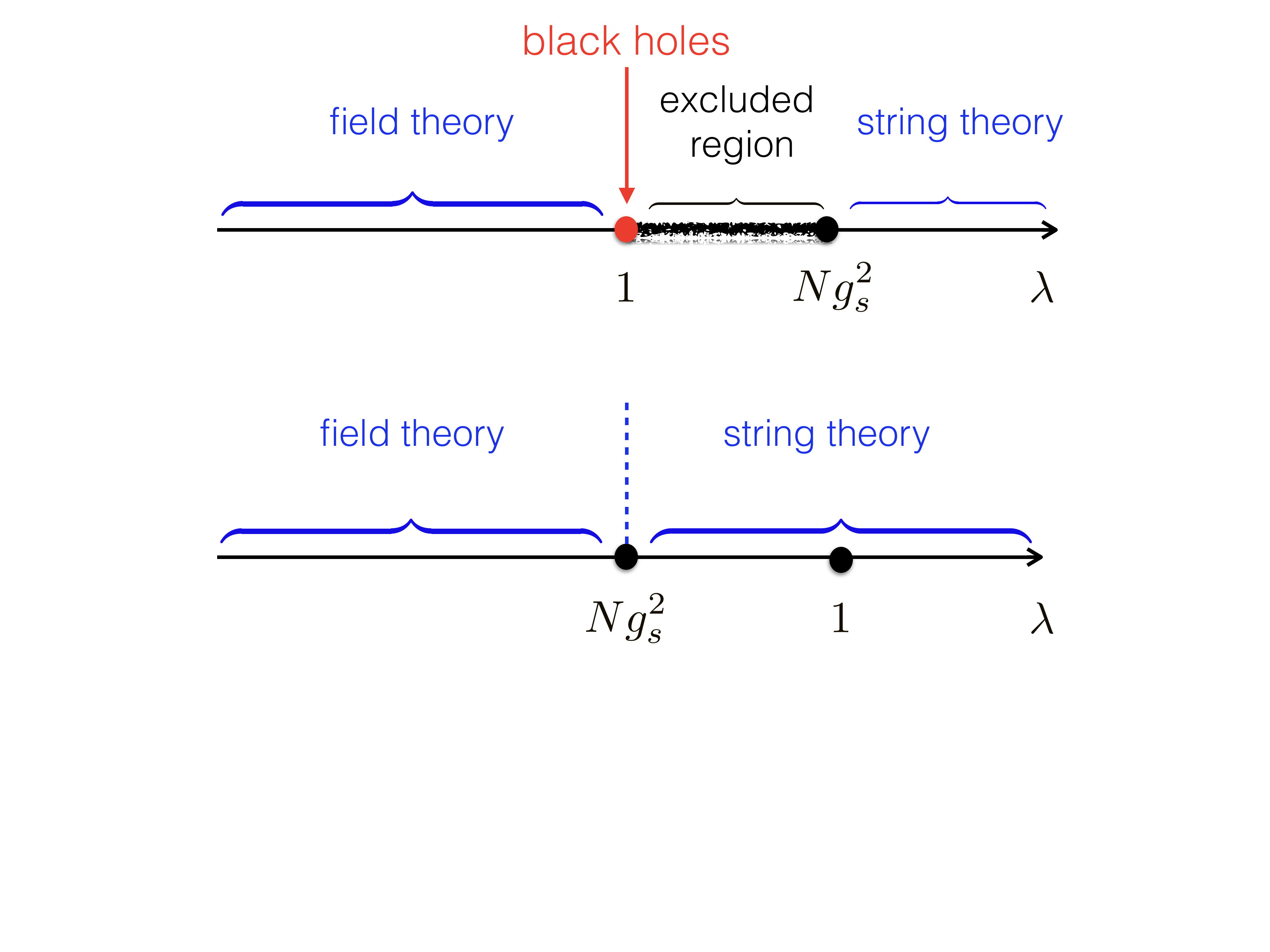}
\vskip-3cm
\caption{Perturbative and non-perturbative regimes as variation of $\lambda$.}
\end{figure}

\noindent 
  
 In the second plot on Fig. 3,  the choice of parameters is such that perturbative string theory and field theory 
 amplitudes crossover without violating unitarity for any $\lambda$.  Of course, for fixed $g_s$ and $L_P$, if we allow $N$ to grow, the situation of the first plot is sooner or later achieved. In other words, for sufficiently high $s$, there is always a window of $\lambda$ for which some perturbatively-allowed  would-be final states violate unitarity.  
 The unitarity in this window is only restored by classicalization,  which excludes it from the Hilbert space due to non-perturbative corpuscular physics of the $N$-graviton system.

      \subsection{Outline}
   
   This work is organized as follows. In the next section we will specify some of the technical steps concerning the  computation of the gravitational scattering amplitudes  for a large number of gravitons in the final state in a specific high energy regime  called classicalization regime.
   Together with other high energy limits the classicalization regime will be defined in section 3. In section 4 we will present the calculation of the gravitational scattering amplitude exhibiting  the details to determine the on--shell scattering amplitude and derive in this way for a large number of $N$ an explicit expression for the transition probability of two particles into $N-2$.   In section 5 an analogous computation is presented for the case of $N$--point open and closed string scattering amplitudes in the high energy limit of classicalization. In particular, we show
   how the relevant combinatorial factor can be derived by using the methods of scattering equations yielding the correct result for the field theory factor appearing in the previous section.
   In the remaining sections we will provide the interpretation of the results for the scattering amplitudes in the light of the  
   corpuscular picture of black holes together with the idea of classicalization. More concretely, in section 6 we extend the discussion of sections 1.2 and 1.3 by explaining
   in which way the full gravitational scattering  amplitude is built as an overlap between the perturbative $N$-graviton amplitude, calculated in sections 4 and 5,  and the non-perturbative projection between the $N$-graviton state and the black hole state, which is provided by the entropy factor $e^N$. This discussion about the 
   perturbative insights into non-perturbative physics is continued in section 7, where
   we also compare the gravitational case with the scalar $\phi^4$ theory. Finally, section 8 contains the outlook of the paper, including also a speculation about the planar limit
   of gauge theories with a large number of colors $N_c$ and the limit of a large number $N$  of gravitons, considered in this paper.

 \section{Recap of technical steps}

  As discussed above,  the basic idea of this paper is to describe  some key aspects of classicalization and black hole formation in the light of high-energy scattering amplitudes
 with a large number~$N$ of soft, elementary quanta in the final state. As already indicated, the phenomenon of classicalization implies a particular high--energy limit of the
 corresponding $N$-particle scattering amplitudes. Specifically we shall analyze the tree-level scattering in the kinematics of $2 \rightarrow (N-2)$ particles, with $N$ being arbitrarily large 
 and with the high center of mass energy $\sqrt {s}$ uniformly distributed over the $N-2$ particles in the final state. We will call this particular kinematical limit the``classicalization" or Eikonal Regge limit.
\begin{figure}[h!]
\centering
\includegraphics[scale=0.4]{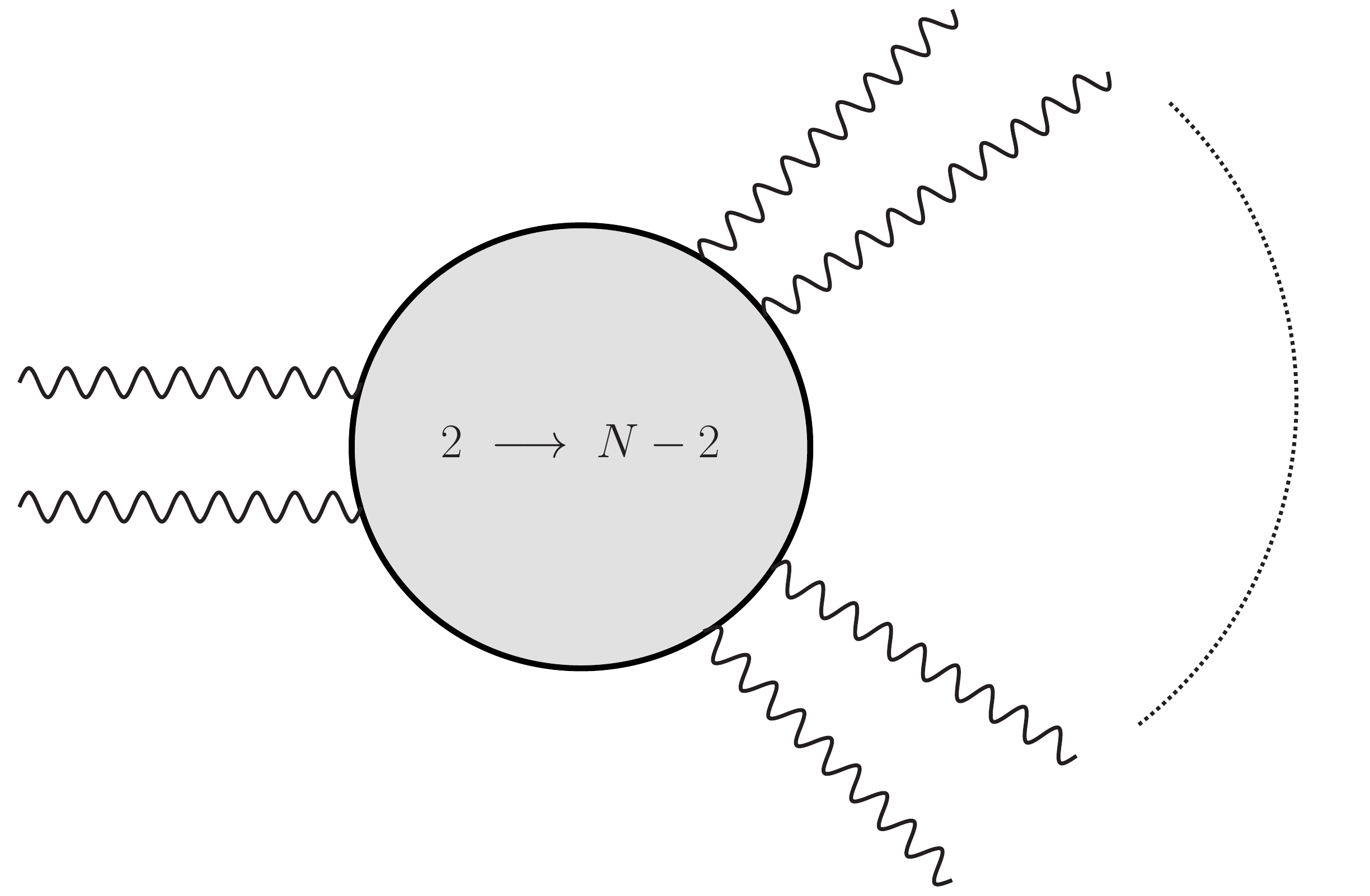}
\caption{Tree level scattering of $2$ into $N-2$ particles. The blob can be thought of as the sum over all Feynman diagrams at tree level.}
\end{figure}

Since it is quite involved to compute the large $N$ field theory amplitudes via standard Feynman diagrams we will employ novel methods in amplitude technology called on-shell methods which were 
developed over the recent years. See \cite{Elvang:2013cua} and references therein for an overview of the vast progress. In addition, we make profit of deriving tree--level amplitudes by the scattering equations \cite{Cachazo}. Concretely, we shall perform the following computations with the following main new results:

\begin{itemize}

%

\item Field theory gravity amplitudes  in the classicalization regime:
We first compute the $2 \rightarrow (N-2)$ graviton amplitudes for arbitrary $N$ in the Eikonal Regge high--energy kinematics. In order to derive these amplitudes we
use a version of the KLT relations for so-called maximally helicity violating (MHV) graviton amplitudes. These are amplitudes with two negative helicity gravitons and the rest positive.
The scattering equations allow us to fix the combinatorial factors of these amplitudes.
From these amplitudes we extract key information about the underlying unitarization mechanism, based on the dominance of this kinematics, as well as on the perturbative suppression factors. These perturbative results provide a strong support both to the physics picture of unitarization by black hole formation as well as to the microscopic picture of black holes as bound states of gravitons. This picture is completed once we superimpose these perturbative results with 
non--perturbative information derived from the many body physics of graviton condensates.

\item Secondly we shall compute the high--energy open/closed string tree level scattering amplitudes for arbitrary (large) number of external legs.
Furthermore we will compare the string amplitudes with field theory amplitudes and discuss the classicalization regime in both cases.  In particular, for fixed $\sqrt{s}$, the two agree for sufficiently large $N$. 
However, depending on the value of the string coupling,  intermediate domains of $N$ are possible 
when either perturbative stringy effects as Reggeization or non-perturbative field theory black hole regimes dominate and exclude 
certain regions. 
 One generic observation is that the regions that are not unitary in perturbative treatment are cut-out by 
 non-perturbative corpuscular black hole physics. On the basis of the concrete form of the string amplitudes we shall make some remarks about some hidden color kinematics duality that at the threshold of black hole formation appears to be reminiscent of the well known gauge gravity duality.


\end{itemize}

\begin{figure}[h!]
\centering
\includegraphics[scale=0.4]{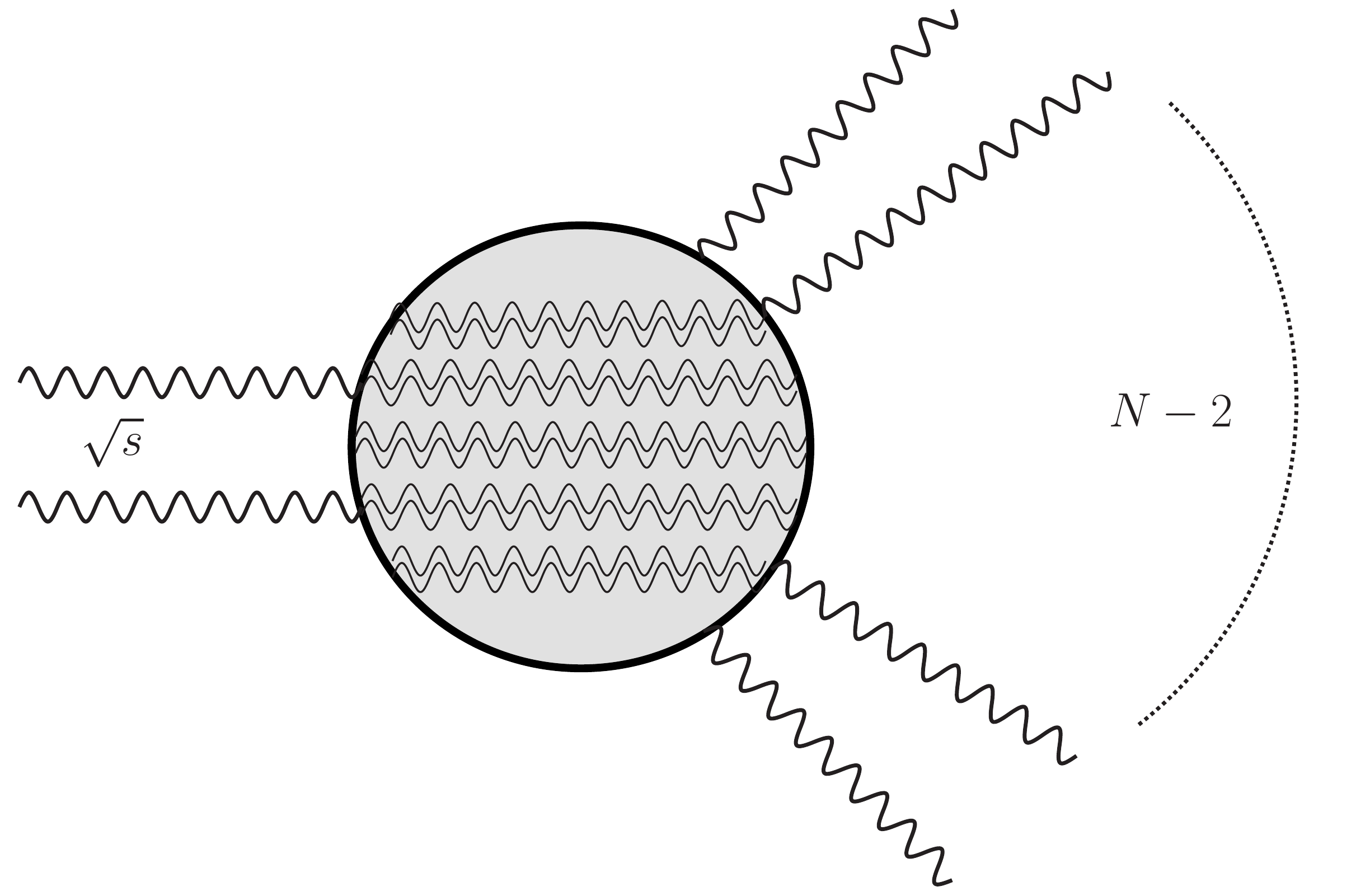}
\caption{Production of a black hole and decay into $N-2$ soft quanta each with momenta $\sim \frac{\sqrt{s}}{N-2}$. The circle with the wiggly double lines depicts the Bose-Einstein condensate nature of the black hole.}
\end{figure}

\section{High--energy kinematical regimes}

In this section we first review the various high--energy limits and their relevance. Generically, for $N$ particle scattering amplitudes
there are $\h N(N-3)$ kinematic invariants
\be\label{kininv}
 s_{ij}\equiv s_{i,j}= (k_i+k_j)^2=2\ k_ik_j\ .
\ee
However, the number of independent invariants depends on the number of space--time dimensions $D$
under consideration. The number of independent Mandelstam variables will be reduced due to Gram determinant relations \cite{asribekov}. Depending on the number of dimensions $D$ and particles $N$, the number of these constraints is given by
\be
\#{\text{constraints}}= \frac{1}{2}(N-D)(N-D-1)
\ee
reducing the number of independent invariants to
\be
\# \{s_{ij}^{indep}\}=N(D-1)-\frac{D(D+1)}{2}.
\ee
In the sequel, however,  we shall not be concerned with this issue and our results are independent of this number.

\subsection{High--energy limits}

For the high--energy limits of the four--point amplitude there are two regions of interest, which 
in this subsection we will review shortly.
The kinematic invariants \req{kininv} for $N=4$ are given by
\be\label{Mandel4}
 s=(k_1+k_2)^2\ \ \ ,\ \ \ t=(k_1+k_3)^2\ \ \ ,\ \ \  u=(k_1+k_4)^2\ ,
\ee 
with $s+t+u=0$.
In the four--point scattering case we have the following relations 
\bea\label{anglestu}
 s&=&\ds{-E^2\ ,}\\[3mm]
 t&=&\ds{E^2\ \sin^2\fc{\theta}{2}\ ,}\\[3mm]
 u&=&\ds{E^2\ \cos^2\fc{\theta}{2}\ ,}
\eea
with $E$ the center--of--mass energy and $\theta$ being the angle between the external momenta $k_1$ and 
$k_3$ (center--of--mass scattering angle).

\subsubsection*{Regge limit}
The Regge limit (also known as small fixed angle regime) is defined as 
  \begin{equation}
  s\gg|t|\gg\Lambda \quad \text{with}\quad \left|\frac{s}{t}\right|\rightarrow\infty,
  \end{equation}
where $\Lambda$ is some scale (usually that of QCD). In this regime
scattering amplitudes of Yang-Mills and gravity field theory
amplitudes exhibit a power-like behavior $\sim s^{\alpha(t)}$ with the
exponent usually called the \textit{Regge slope}. In the usual
treatment, this slope is larger than unity meaning that the amplitude is
not unitary at high energies. In order the unitarize the high-energy behavior of the amplitudes in e.g.\ Yang-Mills it was found that one has to take into account so-called multi--pomeron exchanges which are basically resumed all-loop information of certain ladder-type diagrams. In what follows we shall not touch this unitarization problem appearing in multi Regge kinematics.

\subsubsection*{Hard scattering limit}
The hard scattering limit (fixed  finite angle regime) is the high--energy
domain (ultra high--energy limit) where 
all kinematic invariants become large while their ratios remain fixed. 
In the four--point scattering case  this limit  is defined by:
 \begin{equation}
 \begin{split}\label{HardLimit}
   s, t\rightarrow \infty \quad \text{with} \quad \left|\frac{ s}{ t}\right|\sim\left|\frac{ s}{ u}\right|\sim\left|\frac{ u}{ t}\right|=\text{fixed}\ .
 \end{split}
 \end{equation}
 
\subsection{High--energy limit of four--point field--theory amplitudes}

For the  scattering $p_1+p_2\ra p_3+p_4$ of particles of different mass the differential cross section in the CM frame  is given by a sum over all spins or helicities of scattering subamplitudes $A(1,2,3,4)$
\be\label{CMcrosssection}
\lf.\fc{d\sigma}{d\Omega}\ri|_{CM}=\fc{1}{64\pi^2 E^2}\ \fc{|\vec p_3|}{|\vec p_1|}\ 
\sum_{helicities}|A(1,2,3,4)|^2\ ,
\ee
with $E$ the CM energy. For $e^+e^-\ra \mu^+\mu^-$ the sum over the matrix elements
becomes ($p_{ij}=2p_ip_j$)
\be
\sum_{helicities}|A(1_e,2_e,3_\mu,4_\mu)|^2=\fc{2e^4}{(p_1+p_2)^2}\ \lf[\ p_{13}\ p_{24}+
p_{14}\ p_{23}+2m_\mu^2\ p_{12}+2m_e^2\ p_{34}+8\ m_e^2m_\mu^2\ \ri]\ ,
\ee
with the electron $m_e$ and muon mass $m_\mu$.
In the ultra high--energy limit $m_e,m_\mu=0$, with $|\vec p_1|=|\vec p_3|=\h E$ the cross section \req{CMcrosssection} becomes \cite{Dixon:2013uaa}
\be\label{cmcrosssection1}
\lf.\fc{d\sigma}{d\Omega}\ri|_{CM}=\fc{e^4}{32\pi^2 E^2}\ \fc{t^2+u^2}{s^2}=\fc{e^4}{64\pi^2 E^2}\ \ (1+\cos^2\theta)
\ee
in terms of the quantities \req{anglestu}. 
On the other hand, Rutherford scattering $e^-p^+\ra e^-p^+$ is obtained by using the corresponding
$t$--channel matrix element
\be\label{Rutherford}
\sum_{helicities}|A(1_e,3_p,2_e,4_p)|^2=\fc{2e^4}{(p_1-p_3)^2}\ \lf[\ p_{14}\ p_{23}+
p_{12}\ p_{34}-2m_p^2\ p_{13}+2m_e^2\ p_{24}+8\ m_e^2m_p^2\ \ri]\ ,
\ee
with the proton mass $m_p$. Using \req{Rutherford} and evaluating the corresponding differential cross section for elastic scattering (with $E\sim m_p$) gives the famous Rutherford scattering formula\footnote{Alternatively, in the non--relativistic limit 
the  elastic cross section $\lf.\fc{d\sigma}{d\Omega}\ri|_{CM}= \fc{1}{64\pi^2m_p^2}|A|^2$
can be approximated by the $t$--channel matrix element $A=(-2em_p)\fc{1}{t}(2em_e)$ also 
yielding \req{cmcrosssection2} \cite{Schwartz}.}
\be\label{cmcrosssection2}
\lf.\fc{d\sigma}{d\Omega}\ri|_{CM}=\fc{e^4}{4\pi^2}\ \fc{m_e^2}{t^2}
\ee
in the non--relativistic limit (Born approximation) $m_e^2+p^2\sim m_e^2$ with  $|\vec p_1|=|\vec p_3|=p$.
On the other hand, in the ultra high--energy limit $m_e^2+p^2\sim p^2$, i.e. $m_e\sim0$
and $m_p\ra\infty$ in terms of the quantities \req{anglestu} we have:
\be\label{cmcrosssection3}
\lf.\fc{d\sigma}{d\Omega}\ri|_{CM}=\fc{e^4}{64\pi^2}\ \fc{u}{t^2}\ .
\ee
Obviously, in \req{cmcrosssection2} and \req{cmcrosssection3} the propagator term $1/t^2$
dominates the high--energy behavior of the cross sections.

 \subsection{Eikonal constraints and high--energy limits}
 
 A special region of the space of kinematic invariants \req{kininv} describes the so--called Eikonal constraints.
In this limit two external momenta say $k_1$ and $k_N$ are singled out and kinematic invariants \req{kininv} 
involving neither one of these two momenta nor two non--adjacent momenta are chosen  
to vanish. More precisely, the constraints on
$\h(N-3)(N-4)$ kinematic invariants \req{kininv} is
\be\label{VAR1}
 s_{ij}=0\ ,\ \ \ i=2,\ldots,N-3,\ \ \ i+2\leq j \leq N-1\ ,\\
\ee
while the remaining $2(N-3)$ invariants
\bea\label{VAR2}
s_{1j}&\neq& 0\ ,\ \ \ j=2,\ldots,N\ \ \ ,\ \ \  s_{l,N}\neq 0\ ,\ \ \ l=2,\ldots,N-1\\
 s_{i,i+1}&\neq& 0\ ,\ \ \ i=2,\ldots,N-2
\eea
are left free. E.g. we have:
\bea
N=5&:& s_{24}=0\ ,\\
N=6&:& s_{24}=0,\ s_{25}=0\ , s_{35}=0\ ,\\
N=7&:& s_{24}=0,\  s_{25}=0\ , s_{26}=0,\  s_{35}=0,\  s_{36}=0,\ s_{46}=0\ ,\\
&&\vdots
\eea

As we shall see in section 5 in this Eikonal limit the gauge and gravitational superstring 
amplitudes assume a form which is suited to study properties known from field--theory
amplitudes in the large complex momentum limit \cite{ArkaniHamed:2008yf}.
The latter gives a relation between BCFW relations and the pomeron vertex in string theory
\cite{Brower:2006ea}, cf. also \cite{Cheung:2010vn,Boels:2010bv}. BCFW shifts in string amplitudes have also been studied in \cite{Boels:2008fc}.

\subsubsection*{Eikonal hard scattering limit}

For \req{VAR1} in the hard scattering limit we consider  the limit
$ s\ra\infty$ for the non--vanishing invariants\footnote{We can always find finite parameters to meet these conditions, e.g. for $N=5$ we may choose $s_{12}=s_{45}=-\h s,\ s_{23}=s_{34}=\fc{3}{2}s$ and $s_{51}=s_{23}+s_{34}=3s$.} \req{VAR2}:
\bea\label{VAR3}
 s_{1j}&\sim&  -s\ ,\ \ \ j=2,\ldots,N-1\ \ \ ,\ \ \  s_{l,N}\sim  -s\ ,\ \ \ l=2,\ldots,N-1\ ,\\
 s_{i,i+1}&\sim &  s\ ,\ \ \ i=2,\ldots,N-2\ \ \ ,\ \ \  s_{1N}\sim s\ .
\eea

\subsubsection*{Eikonal Regge limit}

For \req{VAR1} the non--vanishing invariants \req{VAR2} can be parameterized as follows
\begin{equation}\label{ER}
 s_{ij}=(k_i+k_j)^2\sim\begin{cases}
        s\ , \quad i,j\in\{1,N\}\ ,\\
        -\eps\ s\ , \quad i\in\{1,N\}\ ,\ j\notin\{1,N\}\ ,\\
       \eps^2\ s\ ,\quad  i,j\notin\{1,N\}\ ,
        \end{cases}
\end{equation}
with some $ s$ and $\eps$. The (adjacent) Eikonal Regge limit is obtained 
for  small $\eps$ and $s\ra\infty$.
This limit corresponds to a regime, where one subset of momenta (the adjacent momenta $k_1$ and $k_N$) is much greater than a given scale 
$\eps$, while the other subset (all remaining momenta $k_i,\ i\neq 1,N$) is negligible compared to this scale, i.e. $k_i\sim \eps  s^{1/2}$ and $k_1,k_N\sim 
s^{1/2}$.

\subsection{Classicalization high--energy limit}

This is the limit in which we want to analyze the classicalization behavior of the scattering amplitudes. 
This behavior should manifest itself by preferring amplitudes with a greater number of external legs (and vice versa suppressing amplitudes with smaller number of external legs). 
 Take particles $1$ and $N$ to be incoming with center of mass energy $ s:= s_{1N}=(k_1+k_N)^2$ so that their momenta will be proportional to $\sqrt{s}/2$. Accordingly, we define the other $N-2$ particles to be outgoing with momenta proportional to $-\sqrt{ s}/(N-2)$. 
This kinematical choice will lead to a particular scaling of momentum invariants given by
  \begin{equation}
  s_{ij}=(k_i+k_j)^2\sim\begin{cases}\label{DvaliRegime}
       \ds{ s\ , \quad i,j\in\{1,N\}\ ,}\\[3mm]
       \ds{ -\frac{s}{N-2}\ , \quad i\in\{1,N\}\ ,\ j\notin\{1,N\}\ ,}\\[3mm]
       \ds{ \frac{s}{(N-2)^2}\ ,\quad  i,j\notin\{1,N\}\ .}
        \end{cases}
\end{equation}
 Note that \eqref{ER} and \eqref{DvaliRegime} are closely related by identifying
\be\label{eps}
\eps = \frac{1}{N-2}\ .
\ee
The kinematical configurations \req{ER} and \req{DvaliRegime} of the $N$--point amplitude are depicted in Fig.~6.
\begin{figure}[H]
\centering
\includegraphics[scale=0.35]{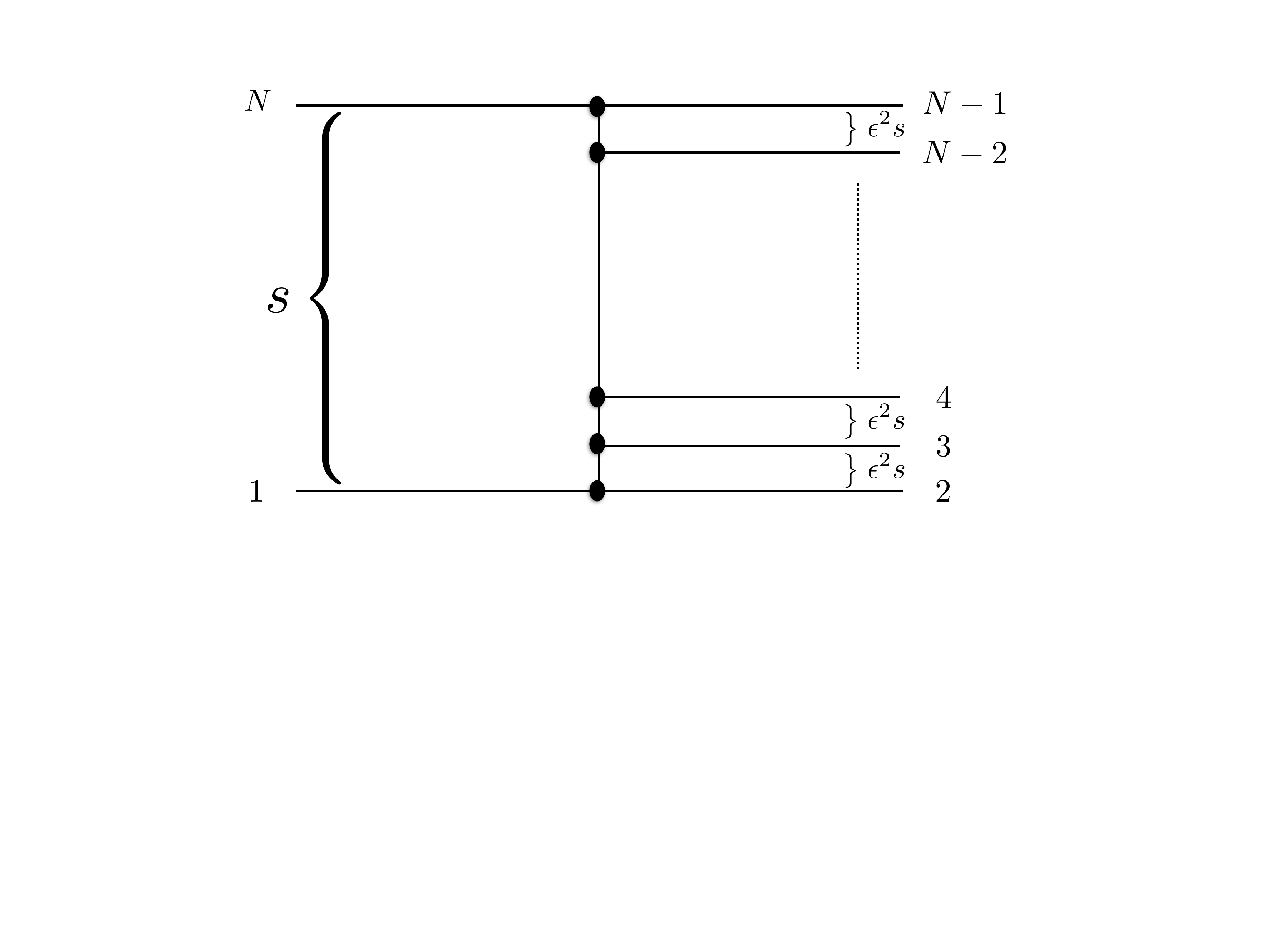}
\vskip-4cm
\caption{Kinematical configuration of the $N$--point amplitude.}
\end{figure}

\section{Field theory perspective}

\subsection*{Large $\bm{N}$ field theory amplitudes in the high--energy classicalization limit}

In this section the high--energy behaviour of field theory scattering amplitudes of pure Yang-Mills theory and pure gravity will be investigated. It will be examined whether one can find hints for classicalization in the high--energy behaviour of these amplitudes. 

The starting point for the field theory computations in gravity are the Kawai-Lewellen-Tye (KLT) relations. These relations express an $N$--point graviton scattering amplitude at tree level 
$\mathcal{M}_{FT}(1,\ldots,N)$ in terms of sums of products of Yang-Mills $N$--point tree amplitudes $A_{YM}$. This was first proven in string theory 
by exploiting the monodromy properties of the closed string world--sheet \cite{Kawai:1985xq} and  later cast into compact form in \cite{Bern:1998sv,BjerrumBohr:2010hn}. Along these lines the $N$-graviton scattering \textit{field} theory tree amplitude becomes
  \begin{equation}\begin{split}\label{BohrKLT}
   \mathcal{M}_{FT}(1,\ldots,N)&=(-1)^{N-3}\ \kappa^{N-2}\sum_{\sigma,\gamma\in S_{N-3}}A_{YM}(1,\sigma(2,\dots,N-2),N-1,N)\\
   &\times S[\gamma(2,\dots,N-2)|\sigma(2,\dots,N-2)]_{N-1}\ A_{YM}(1,N-1,\gamma(2,\dots,N-2),N)\ ,
  \end{split}\end{equation}
  where the sum runs over the permutations of the $N-3$ elements of the sets $\gamma$ and $\sigma$. 
Above $\kappa$ is the gravitational coupling constant with $\kappa^2=16\pi G_N$.
The function $S[\dots|\dots]$ is called \textit{momentum kernel} or \textit{KLT kernel} and is defined via the Mandelstam variables \req{kininv} as
  \begin{equation}\label{Kernel}
   S[{i_1,\dots,i_k}|{j_1,\dots,j_k}]_P=\prod_{t=1}^k\left(s_{i_t,P}+\sum_{q>t}^k\theta(i_t,i_q)\ 
 s_{i_t,i_q}\right).
  \end{equation}
Here, $P$ is a lightlike reference momentum. The function $\theta$ is defined such that
  \begin{equation}
   \theta(i_a,i_b)=\begin{cases} 0, & \text{if $i_a$ sequentially comes before $i_b$ in $\{j_1,...j_k\}$}\\
		   1, & \text{else}.
                   \end{cases}
  \end{equation}
Note that while the gravity amplitude above is only manifestly $S_{N-3}$ invariant, in fact it can be shown to be completely permutation symmetric.

\subsubsection*{Hard scattering limit}
In the hard scattering limit, which was defined in equation \eqref{HardLimit} for four points, the behavior of field theory amplitudes is well known. Define $s=(k_1+k_N)^2$ to be the center of mass energy. By counting mass dimensions one finds that an $N$-point Yang-Mills amplitude behaves as
  \begin{equation}
  \label{FTbev1}
   A_{YM} \sim s^{\frac{4-N}{2}},
  \end{equation}
i.e.\ it displays a power-fall off such that the amplitude decreases as the energy increases. In this way unitarity at tree level will not be violated. 
In contrast, an $N$--point gravity amplitude displays the following behavior
  \begin{equation} \label{FTbev2}
  \mathcal{M}_{FT} \sim \kappa^{N-2}\   s.
  \end{equation}
In other words, the $N$-graviton amplitude grows monotonically as $s$ increases.
Alternatively, the high-energy behavior of the graviton amplitude can also be seen from the KLT formula 
\req{BohrKLT}  taking into account that each of the entries of the momentum kernel behave as
  \begin{equation}
 S[\sigma(2,...,N-2)|\gamma(2,...,N-2)]_P \sim  s^{N-3}
  \end{equation}
in the hard scattering limit.

\subsubsection*{Classicalization high-energy limit}

Next, let us analyze whether we can find indications of classicalization in the high--energy limit of graviton scattering amplitudes utilizing \eqref{BohrKLT}. The classicalization high--energy limit was already defined in eq. (\ref{DvaliRegime}).

Consider first Yang-Mills amplitudes in the high--energy limit as they are a building block of gravity amplitudes and so will be needed later. For simplicity, let us restrict ourselves to the case of MHV amplitudes, i.e\ amplitudes with two particles of negative helicity and the other ones with positive helicity. In this case, the Yang-Mills amplitudes take a particularly simple form in four dimensions. In standard spinor helicity notation, it is given by \cite{Parke:1986gb}
  \begin{equation}\label{MHV}
   A_{YM}(1^+,...,i^-,...,j^-,...,N^+)=\frac{\langle i\;j\rangle^4}{\langle 1\;2 \rangle\langle2\;3\rangle...\langle N-1\;N\rangle\langle N\;1\rangle}.
  \end{equation}
The spinor helicity brackets are basically square roots of Mandelstam invariants and their precise relation is given by \cite{Dixon:1996wi}
  \begin{equation}\label{BraToMa}
   \langle i j \rangle=\sqrt{| s_{ij}|}e^{i\phi_{ij}} 
  \end{equation}
with
  \begin{equation}\label{phase}
   \cos(\phi_{ij})=\frac{k_i^1k_j^+-k_j^1k_i^+}{\sqrt{| s_{ij}|k_i^+k_j^+}},\quad \sin(\phi_{ij})=\frac{k_i^2k_j^+-k_j^2k_i^+}{\sqrt{| s_{ij}|k_i^+k_j^+}},\quad \text{and} \quad k^{\pm}=k^0\pm k^3.
  \end{equation}
In the classicalization kinematics region one can straight-forwardly find the scaling of the Yang-Mills amplitude by applying \eqref{BraToMa} and \eqref{DvaliRegime} to \eqref{MHV}. Notice that one has to distinguish three cases: either both incoming particles have negative (positive) helicity or one particle might be positive and the other negative. One finds for these three cases (suppressing the absolute value)
  \begin{equation}\label{ThreeCases}
   A_{YM}(i^-,j^-)\sim  s^{\frac{4-N}{2}}\ f(\phi)\ \times\begin{cases}(N-2)^{N-2}\ ,\quad i,j\in\{1,N\}\ ,\\
	(N-2)^{N-4}\ , \quad i\in\{1,N\}\ ,\ j\notin\{1,N\}\ ,\\
(N-2)^{N-6}\ , \quad  i,j\notin\{1,N\}\ ,
                \end{cases}
  \end{equation}
with $f(\phi)$ shorthand for the phase-factors. In general $f(\phi)$ will be a very complicated function. The difference in scaling above can be easily understood: it originates in the numerator of the MHV expression \req{MHV}. The high--energy scaling of $\langle i\;j\rangle^4$ depends on which particles are chosen to have negative helicity. The denominator, on the other hand, just encodes the pole structure of the amplitude which is independent of helicities. In other words, the denominator scales in the same way in all three cases. Note that the phase factors \eqref{phase} do not scale with $N-2$ since the scaling cancels.

Let us now turn to gravity amplitudes. To use the KLT relation \eqref{BohrKLT} we need to know how the momentum kernel scales in this regime: its entries roughly scale as
  \begin{equation}
  S[\gamma(2,...,N-2),\sigma(2,...,N-2)]_{N-1}\sim \left(\frac{ s}{(N-2)^2}\right)^{N-3} 
  \end{equation}
as all particle labels involved in the momentum kernel belong to outgoing particles. Combing this with the scaling of the Yang-Mills amplitudes \eqref{ThreeCases} via \eqref{BohrKLT}, leads to the following scaling for the gravity MHV amplitude 
  \begin{equation}
  \mathcal{M}_{FT}(i^-,j^-)\sim   \kappa^{N-2}\ \tilde{C}(N)\ s\ \times\begin{cases} (N-2)^{2}\ , \quad i,j\in\{1,N\}\ ,\\
					      (N-2)^{-2}\ , \quad i\in\{1,N\}\ ,\ j\notin\{1,N\}\ ,\\
                                            (N-2)^{-6}\ , \quad  i,j\notin\{1,N\}\ ,
                              \end{cases}
  \end{equation}
where the function $\tilde{C}(N)$ is a complicated double sum over the phase factors arising when one rewrites the spinor brackets in terms of Mandelstam invariants and sums over the different permutations in the KLT sum. Unfortunately, it is very involved to evaluate this sum by using the techniques under consideration. However, fortunately as we shall see in subsection 5.3.3, the factor can be computed in a straightforward way by making profit of the scattering equations \cite{Cachazo}. It will be shown that (see equation \eqref{ESTI})
\be\label{derFactor}
\tilde{C}(N)=(N-1)!
\ee
Hence, in the classicalization regime the scaling of the gravity field theory amplitude at tree level should be given by
  \begin{equation}\label{Nscaling}
  \mathcal{M}_{FT}(i^-,j^-)\sim   \kappa^{N-2}\ (N-1)!\  s\ \times\begin{cases} (N-2)^{2}\ , \quad i,j\in\{1,N\}\ ,\\
					      (N-2)^{-2}\ , \quad i\in\{1,N\}\ ,\ j\notin\{1,N\}\ ,\\
                                            (N-2)^{-6}\ , \quad  i,j\notin\{1,N\}\ .
                              \end{cases}
  \end{equation}
Note that we could have chosen \textit{any} two particles to be
incoming in the analysis above. The only difference would be in the
scaling of the momentum kernel: it would scale more complicated as a
function of $N$ since the kinematic invariants would not all scale homogeneously like $ \tfrac{s}{N-2}$. Naively, the behavior of the gravity amplitude would then also be more complicated as a consequence. However, due to Bose symmetry the overall result is independent of the choice of incoming momenta and the seemingly more complex scaling will cancel in the sum over terms in \eqref{BohrKLT} giving back equation \eqref{Nscaling}.

The next step is to go from the on-shell scattering amplitude above to the physical (dimensionless) transition probability of two particles scattering into $N-2$, i.e. to $ |\langle 2|S|N-2\rangle|^2$ (cf. equation (1.1)). In order to do so, one has to multiply $\mathcal{M}_{FT}$ by the values of the outgoing and incoming momenta and take into account that the final states are identical. This amounts to dividing out by a factor of $(N-2)!$. Doing so one arrives at (indices suppressed)
\be 
|\langle 2|S|N-2\rangle|^2=\frac{1}{(N-2)!}\left(\prod_{i=2}^{N-1}p_i \,p_1p_N\ \mathcal{M}_{FT}\right)^2,
\ee
Finally, by identifying $\kappa$ with $L_P$, taking $N\gg 1 $, and remembering  that $p_{out}\sim \frac{\sqrt{s}}{N}$ and $p_{in}\sim \sqrt{s}$ one arrives at equation (1.1)
\begin{equation}\label{newGR}|\langle 2|S|N\rangle|^2 \, \sim \,  \left ({L_P^2  s \over N^2} \right )^{N} \, N!.
 \end{equation}
 Note that for large $N$ we do not need to make the distinction between the three different cases anymore as they all scale the same. Moreover, the result \req{newGR} holds for both  MHV and NMHV  scattering since in subsection 5.3.3 for the derivation of the factor \req{derFactor} 
no specific helicity configuration is assumed.

 As already advertised in the introduction this physical amplitude starts to unitarize for $N$ given by $s =N M_P^2$. Moreover the amplitude at this kinematical unitarity threshold is suppressed by $e^{-N}$ anticipating at this perturbative level exactly the suppression factor that can be compensated by the entropy of a black hole with mass $M=\sqrt{ s} = \sqrt{N}M_P$.

\section{String theory perspective}
\def\Af{{\frak A}}
\def\Ac{{\mathcal A}}\def\Oc{{\mathcal O}}
\def\Mc{{\mathcal M}}
\def\si{\sigma}
\def\IZ{{\bf Z}}
\def\sv{{\rm sv}}
\def\br{{\langle}}
\def\ke{{\rangle}}

The high--energy behavior (i.e. energies much larger than the string scale $M_s$) of perturbative string amplitudes is rather different than that of field--theory amplitudes.
While the high--energy behavior in field--theory \req{cmcrosssection2} furnishes a power fall--off  behavior (in the kinematic invariant~$t$) string theory exhibits an exponential fall--off.
This opens the possibility to investigate the unitarity properties of the field theory amplitudes at high (or even trans--Planckian) energies within the framework of perturbative string amplitudes. The latter take into account effects from higher spin and black hole states, which may play a crucial role in the unitarization of the amplitude. In fact, higher spin states are vital for the consistency of weakly coupled gravity theories in the tree--level approximation 
with higher derivative corrections at intermediate energies (energies, at which the theory is still   weakly coupled, but sensitive to higher derivative corrections) \cite{Camanho:2014apa}. 
 
In string theory besides the string scale $M_s$, which is related to the string length $L_s$ and  string tension 
$\ap$ as
\be
\ap\sim L_s^2= M_s^{-2}
\ee
there is a string loop expansion parameter $g_s\sim e^{\Phi}$ (or closed string coupling constant $g_{closed}=g_s=\kappa_4/\ap^{1/2}$) controlling 
higher genus string effects with the dilaton field $\Phi$. 
The latter determines the YM coupling $g_{YM}\sim e^{\Phi/2}$, which enters as the open string
coupling $g_{open}=g_{YM}$.
The relation between the Planck and string mass is given by $M_P= g_s^{-1}M_s$, i.e.
\be\label{Planck}
L_P=g_s\ L_s\ ,
\ee 
which in turn implies:
\be\label{OPCL}
g_{closed}=g_{open}^2\ .
\ee
Hence, a small string loop expansion parameter $g_s$ corresponds to:
\be
g_s^{-2}\sim\fc{M_P^2}{M_s^2}=\fc{L_s^2}{L_P^2}\gg1\ .
\ee
On the other hand, energies for which tree amplitudes are large correspond to $ sL_P^2 \gg 1$.
Hence, the legitimate kinematical regime to study high--energy string tree--level  scattering is:
\be
\ap s\gg \fc{L_s^2}{L_P^2}\gg 1\ .
\ee

The string theory scattering at fixed angles and large energy is determined by a classical solution, i.e. the high--energy scattering in string theory becomes semi--classical:
the two--dimensional string world--sheet stretches at long distance and the classical solution minimizes its area. The world--sheet string integrals are dominated by saddle points.

The high--energy  fixed--angle behavior of open string tree--level scattering was first investigated by Veneziano \cite{Veneziano}. 
The high--energy behavior of 
four--point string scattering was then thoroughly analyzed by Gross, Mende and Manes \cite{Gross:1987kza}, see also  \cite{Amati:1987wq} for some complementary work. In this section for  tree--level
and a specific kinematical region (specified in section 3) 
we shall generalize these results  to an arbitrary number of external string states. 
Further interesting aspects of high--energy behavior in string theory have been discussed in  \cite{D'Appollonio:2010ae}.

Beyond the Born approximation (cf. e.g. eq. \req{cmcrosssection2}) for smaller impact parameter (larger $t$) the Eikonal scattering regime is reached where ladder diagrams (and crossed ladder diagrams) in the $s$--channel become
important\footnote{Some interesting connections between ladder diagrams and the picture of graviton Bose--Einstein condensates have recently been presented in \cite{Kuhnel:2014xga}.} \cite{Giddings:2011xs}. In field theory the latter can be derived from or matched to available perturbative higher--loop supergravity computations \cite{Giddings:2010pp}.
On the other hand, in string theory it has already been argued in \cite{Gross:1987kza}
that the higher genus high--energy behavior can also be approximated by one 
saddle point showing the  universal exponential behavior \req{highGf}.
Hence, we believe, that our tree--level 
high--energy results may at least qualitatively also describe effects from higher genus 
string world--sheet topologies.

\subsection{High--energy limit of four--point open and closed superstring amplitude}

The color ordered open superstring four--point tree  subamplitude reads
\be\label{startYM}
\Ac(1,2,3,4)=g_{YM}^2\ A_{YM}(1,2,3,4)\ F_4\ ,
\ee
with the SYM subamplitude $A_{YM}(1,2,3,4)$ and  the string form factor:
\be\label{Form4}
F_4=\fc{\Gamma(1+\ap s)\ \Gamma(1+\ap u)}{\Gamma(1+\ap s+\ap u)}\ .
\ee 
The kinematic invariants  are given in  \req{Mandel4}.
In the hard scattering limit, i.e.\ for $\alpha'\rightarrow \infty$ the form factor
 \req{Form4} behaves as
\begin{equation}
   F_4 \sim (2\pi\ap)^{1/2}\ \lf|\fc{su}{t}\ri|^{1/2}\ \exp\left\{\ap\lf(s\;\ln |s|+t\;\ln t+u\;\ln u \ri)\right\}\ ,
   \label{high4}
  \end{equation}
to be contrasted with the corresponding field--theory expression \req{cmcrosssection2}
in the Born approximation.
Eventually, with the fixed--angle parameterization \req{anglestu}  eq. \req{high4} can be cast into:
\be\label{steps}
 F_4 \sim (2\pi)^{1/2}\ \fc{E}{M_{\rm string}} \cot\fc{\theta}{2}\ \exp\left\{\fc{E^2}{M_{\rm string}^2}\ \lf(\sin^2 \fc{\theta}{2}\;\ln\sin^2\fc{\theta}{2}+
 \cos^2 \fc{\theta}{2}\;\ln\cos^2\fc{\theta}{2}\ri)\ri\}\ .
\ee
The expression \req{high4} follows by applying the Laplace method. The latter approximates the integral for a function 
$f$ with a unique global maximum $x_0$ inside the integration region as:
\be\label{Laplace}
\int_a^b g(x)\ \exp\lf\{\ap f(x)\ri\}dx\sim\sqrt{\fc{2\pi}{\ap|f''(x_0)|}}\ 
\lf[g(x_0)+\Oc(\ap^{-1})\ri]\ \exp\lf\{\ap f(x_0)\ri\}\ .
\ee
Then, the result \req{high4} follows from rewriting \req{Form4} as
\be\label{stat4}
F_4 = \ap s\ \int_0^1 dx\ x^{\ap s-1}\ (1-x)^{u}=\ap s\ \int_0^1dx\ 
x^{-1}\ \exp\lf\{\ap s\ln x+\ap u\ln(1-x)\ri\}
\ee
and applying Laplace's method \req{Laplace} to approximate the latter. The stationary point 
$x_0=-\fc{s}{t}$ (with $s<0$) of the integrand of \req{stat4} follows
from solving the equation:
\be\label{seq}
\fc{s}{x}-\fc{u}{1-x}=0\ .
\ee
With this information and the formula \req{Laplace} we arrive at \req{high4}.

Furthermore, the closed--string four--point amplitude describing four graviton scattering 
is given by
\be
\Mc(1,2,3,4)=\kappa^2\  |A_{YM}(1,2,3,4)|^2\ (\ap s)^2\ \int_{\bf C}d^2z\  |z|^{\ap s/2-2}\ |1-z|^{\ap u/2}
\ee
with the SYM subamplitude $A_{YM}(1,2,3,4)$ and gravitational coupling constant $\kappa$.
Performing a saddle point approximation in the complex plane (w.r.t. polar coordinates) yields:
\bea\label{highGf}
\ds{\Mc(1,2,3,4)}&\sim&\ds{\kappa^2\  |A_{YM}(1,2,3,4)|^2}\\[5mm]   
&\times&\ds{4\pi\ap\ \lf|\fc{su}{t}\ri|\ \exp\left\{\fc{\ap}{2}\lf(s\;\ln |s|+t\;\ln t+u\;\ln u \ri)\right\}\ .}
\eea
It is interesting to note, that eq. \req{highGf} essentially is the square of the open string case \req{high4} subject to a rescaling of the string tension $\ap$ as $\ap\ra\ap/4$. Hence, qualitatively there is no difference between the high--energy behavior of the open and closed superstring tree--level amplitude. 
This fact becomes feasible  by the single--valued projection \cite{Stieberger:2013wea,Stieberger:2014hba}, cf. the next subsection.

\subsection[High--energy limit of $N$--point gauge and graviton amplitude]{High--energy limit of $\bm{N}$--point gauge and graviton amplitude}

In this subsection  for specific kinematical regions we shall derive the high--energy behaviour of both open and closed superstring scattering amplitudes for an arbitrary number $N$ of external string states.

A first glance at the high--energy behaviour of the $N$--point open superstring amplitude can be gained by considering the Selberg integral \cite{Selberg}
\bea\label{selberg}
\ds{S_n(\alpha,\beta,\gamma)}&=&\ds{\lf(\prod_{i=1}^{n}\int_0^1dt_i\ri)\ 
\prod_{l=1}^n t_l^{\alpha-1}\ (1-t_l)^{\beta-1}\ \prod_{1\leq i<j\leq n}|t_i-t_j|^{2\gamma}\ ,}
\\[6mm]
&=&\ds{n!\ \prod_{l=0}^{n-1} \fc{\Gamma(\alpha+l\gamma)\ \Gamma(\beta+l\gamma)\ 
\Gamma(\gamma+l\gamma)}{\Gamma(\alpha+\beta+(n+l-1)\gamma)\ \Gamma(\gamma)}\ ,}
\eea
with complex parameters $\alpha,\beta,\gamma$ such that 
$\Re\alpha,\Re\beta>0$ and $\Re\gamma>-{\rm Min}\{\fc{1}{n},\fc{\Re\alpha}{n-1},
\fc{\Re\beta}{n-1}\}$.
The Selberg integral \req{selberg} may be thought as the straightforward  multi--dimensional generalization of \req{stat4} to be suited to describe the $N$--point case.
In fact, for $n=N-3$ and the parameterization
\bea
\alpha&=&\ap\ s_{1i}\ \ \ ,\ \ \ \beta=\ap\ s_{i,N-1},\ i=2,\ldots,N-2\ ,\\[2mm]
2\gamma&=&\ap\ s_{ij}\ \ \ ,\ \ \ 2\leq i<j\leq N-2
\eea
eq. \req{selberg} describes a generic world--sheet disk integral involving $N$ open strings.
The integral \req{selberg} sums up $n!$ iterated real integrals with identical contributions.
The latter corresponds  to $(N-3)!$ (independent) color ordered subamplitudes.
With Stirling's formula \cite{Abramowitz}
\bea\label{Stirling}
\ds{\Gamma(z)}&=&\ds{(2\pi)^{1/2}\ \exp\lf\{\lf(z-\h\ri)\ \ln z-z\ri\}}\\[5mm]
&\times&\ds{ \lf(1+\fc{1}{12z}+\fc{1}{288z^2}-\ldots\ri)\ \ \ ,\ \ \ 
\mbox{for}\  z\ra\infty\ \mbox{in}\ |\arg(z)|<\pi}
\eea
we may easily determine the $\ap\ra\infty$ limit of \req{selberg}.
E.g. for $N=5$, i.e. $n=2$ we find the behaviour:
\bea\label{qualSelberg}
S_2&\sim&\ds{ (2\pi)\ 2^{\h+2\gamma} \lf(\fc{(\alpha+\beta+\gamma)(\alpha+\beta+2\gamma)}{\alpha\beta(\alpha+\gamma)(\beta+\gamma)}\ri)^{1/2}}\\[5mm]
&\times&\ds{\exp\lf\{\fc{\alpha\ln\alpha+\beta\ln\beta+\gamma\ln\gamma+(\alpha+\gamma)\ln(\alpha+\gamma)+(\beta+\gamma)\ln(\beta+\gamma)}{(\alpha+\beta+\gamma)\ln(\alpha+\beta+\gamma)+(\alpha+\beta+2\gamma)\ln(\alpha+\beta+2\gamma)}\ri\}+\Oc(\ap^{-1})\ .}
\eea

\subsubsection[High--energy limit of $N$--point gauge  amplitude]{High--energy limit of $\bm{N}$--point gauge amplitude}

The open superstring $N$--gluon tree--level amplitude $\Af_N$ describing the scattering 
of $N$ gluons
 decomposes into a sum
\be\label{DECO}
\Af_N=\sum_{\Pi\in S_{N}/{\bf Z}_2}
\Tr(T^{a_{\Pi(1)}}\ldots T^{a_{\Pi(N)}})\ \Ac(\Pi(1),\ldots,\Pi(N))
\ee
over color ordered subamplitudes $\Ac(\Pi(1),\ldots,\Pi(N))$ 
supplemented by a group trace
in the adjoint representation. 
The sum runs over all permutations $S_N$ of labels $i=1,\ldots,N$
modulo cyclic permutations ${\bf Z}_2$, which preserve the group trace.
The $(N-3)!$ independent open $N$--point superstring subamplitudes can be cast into the 
compact form \cite{Mafra}
\be\label{OPEN}
\Ac(1,\pi(2,\ldots,N-2),N-1,N)=g_{YM}^{N-2}\ \sum_{\sigma\in S_{N-3}}
F_{\pi\sigma}\ A_{YM}(\si)\ ,\ \ \ \pi\in S_{N-3}\ ,
\ee
with  the $(N-3)!$ (independent) SYM subamplitudes
$A_{YM}(\si):=A_{YM}(1,\si(2,\ldots,N-2),N-1,N),\ \si\in S_{N-3}$ 
and the $(N-3)!\times(N-3)!$ matrix $F$, whose entries $F_{\pi\sigma}$ can be expressed
 as  \cite{Broedel:2013tta}
\be\label{stringBCJ}
F_{\pi\sigma}=(-\ap)^{N-3}\ \sum_{\rho\in S_{N-3}}Z_\pi(\rho)\ S[\rho|\sigma]\ ,
\ee
with  the world--sheet disk integrals specified in eq. \req{Start} and some variant  of the KLT kernel \req{Kernel}
\be\label{KLTKERN}
S[\rho|\si]:=S[\, \rho(2,\ldots,N-2) \, | \, \si(2,\ldots,N-2) \, ] = \prod_{j=2}^{N-2} \Big( \, s_{1,j_\rho} \ + \ \sum_{k=2}^{j-1} \theta(j_\rho,k_\rho) \, s_{j_\rho,k_\rho} \, \Big)\ ,
\ee
with $j_\rho=\rho(j)$ and  $\theta(j_\rho,k_\rho)=1$
if the ordering of the legs $j_\rho,k_\rho$ is the same in both orderings
$\rho(2,\ldots,N-2)$ and $\si(2,\ldots,N-2)$, and zero otherwise.
The matrix entries $F_{\pi\sigma}$ given in eq. \req{stringBCJ} represent generalized Euler integrals integrating to  multiple hypergeometric functions \cite{Mafra}:
\be\label{revol}
F_{\pi\si} = (-\ap )^{N-3}\int\limits_{D(\pi)}  
\lf(\prod_{j=2}^{N-2} dz_j\ri)\  \lf(\prod_{i<l} |z_{il}|^{\ap s_{il}}\ri) \ 
\left\{\ \prod_{k=2}^{N-2}  \sum_{m=1}^{k-1} \fc{ s_{mk} }{z_{mk}} \ \right\}\  .
\ee
Above, the permutations $\si\in S_{N-3}$ act on all indices $\{2,\ldots,N-2\}$ within the curly bracket.
Due to conformal invariance on the world--sheet  we have fixed three of the $N$ world--sheet positions as
\be\label{fix}
z_1=0\ \ \ ,\ \ \ z_{N-1}=1\ \ \ ,\ \ \ z_N=\infty\ ,
\ee
and the remaining $N-3$  positions $z_i$ are integrated along the boundary of the disk subject to the ordering $D(\pi)=\{z_j\in \IR\ |\ z_1<z_{\pi(2)}<\ldots<z_{\pi(N-2)}<z_{N-1}<z_N\}$.
Furthermore, we have $z_{ij}\equiv z_{i,j}=z_i-z_j$.
Integration by parts admits to simplify the  integrand in \req{revol}. As a result the length of  the sum over $m$ becomes shorter for $k > \floor{N/2}$
\be\label{revoli}
F_{\pi\si}= (-\ap )^{N-3} \int\limits_{D(\pi)}  
\lf(\prod_{j=2}^{N-2} dz_j\ri) \ \lf(\prod_{i<l} |z_{il}|^{s_{il}}\ri)\ 
\lf\{\left( \prod_{k=2}^{\floor{N/2}} \  \sum_{m=1}^{k-1} \ \fc{ s_{mk}}{z_{mk}}  \ri)\ 
\lf(\prod_{k=\floor{N/2}+1}^{N-2}  \sum_{n=k+1}^{N-1} \ \fc{ s_{kn}}{z_{kn}}  \right)\ri\}\  ,
\ee
with $\floor{x}$ the integer part of $x$.

In the sequel, in \req{OPEN}, without loss of generality let us concentrate on the canonical color ordering $\pi=1$ describing the string subamplitude $\Ac(1,\ldots,N)$ and work out the  
latter in the Eikonal limit \req{VAR1} and \req{VAR2}.
By applying partial integrations w.r.t. to the world--sheet coordinates it can be evidenced, that 
for the case $\si\neq id$ all functions $F_{1\si}$ have one of the 
invariants \req{VAR1} as prefactor. 
Note, that in a gluon subamplitude with canonical color ordering $\pi=1$ the constraints 
\req{VAR1} do not cause any singularities as the latter would correspond to unphysical  poles. 
So we can safely take the limit \req{VAR1}, i.e. $F_{1\si}=0,\ \si\neq 1$ and only $F_N:=F_{11}$ 
is non--vanishing in the Eikonal limit.
As a consequence in the Eikonal limit the full open superstring subamplitude \req{OPEN}
reduces to one term
\be\label{INFOPEN}
\Ac(1,\ldots,N)=g_{YM}^{N-2}\ F_N\ A_{YM}(1,\ldots,N)\ ,
\ee
with the string form factor $F_N$ given by:
\be\label{FORMF}
F_N= (-\ap )^{N-3}\  
\int\limits_{z_i<z_{i+1}}\lf(\prod_{j=2}^{N-2} dz_j\ri)\  \lf(\prod_{i<l} |z_{il}|^{\ap s_{il}}\ri)\ \fc{s_{12}}{z_{12}}\ 
\lf(\prod_{l=1}^{N-4}\fc{s_{N-l-1,N-l}}{z_{N-l-1,N-l}}\ri)\ .
\ee
In the Eikonal limit the kinematical factor $A_{YM}(1,\ldots,N)$ of the superstring amplitude  \req{INFOPEN}  
is identical in form to that of the corresponding field theory amplitude.
Hence, in the four--dimensional MHV 
case\footnote{According to \cite{Stieberger:2012rq} the rational function in the world--sheet positions has a simple representation in terms of tree diagrams. In the MHV case (with gluon $1$ and $N$ of negative helicity) the $N$--point open superstring subamplitude \req{OPEN} can be expressed as \cite{Stieberger:2012rq,Stieberger:2013hza}
\bea
&&\ds{\Ac(1,\ldots,N)=g_{YM}^{N-2}\ \frac{\vev{1N}^4}{\vev{1,(N-1)}\vev{ (N-1),N}\vev{N1}}}\\[5mm]
&&\hskip0.5cm\times\ds{\int\limits_{z_i<z_{i+1}}\lf(\prod_{j=2}^{N-2} dz_j\ri)  \lf(\prod_{i<l} |z_{il}|^{\ap s_{il}}\ri) \sum_{\si\in S_{N-2}} \fc{1}{z_{12}}\ \prod_{k=2}^{N-2}\frac{\br N|N-1+\ldots+(k+1)|k]}{\br kN\ke} 
\fc{\ap}{z_{k(k+1)}}\ ,}
\label{samp1}
\eea
where the permutations $\si$ act on the set $\{2,\ldots,N-2\}$. In the above form \req{samp1} 
one easily observes the effect of taking the Eikonal limit \req{VAR1}. Any of the $(N-3)$ brackets
$\br N|(N-1)+\ldots+(k+1)|k]$ would vanish for $\si\neq 1$.
Hence the Eikonal limit of \req{samp2} gives
\bea
\ds{\Ac(1,\ldots,N)}&=&\ds{g_{YM}^{N-2}\ \frac{\vev{1N}^4}{\vev{1,(N-1)}\vev{ (N-1),N}\vev{N1}}}\\[5mm]
&\times&\ds{\int\limits_{z_i<z_{i+1}}\lf(\prod_{j=2}^{N-2} dz_j\ri)\  \lf(\prod_{i<l} |z_{il}|^{s_{il}}\ri)\ \fc{1}{z_{12}}\ \prod_{k=2}^{N-2}\frac{\br N|(k+1)|k]}{\br kN\ke}\  \fc{\ap}{z_{k(k+1)}}\ ,}
\label{samp2}
\eea
which can be shown to agree with \req{INFOPEN}  in the MHV case.} 
in eq. \req{INFOPEN} the SYM amplitude factor is given by the  
Parke--Taylor amplitude \req{MHV}.

For the choice of vertex operator positions \req{fix} and  the parameterization
\be\label{FIX}
z_l=\prod_{i=1}^{N-l-1}x_i\ \ \ ,\ \ \ l=2,\ldots,N-2
\ee
the string form factor \req{FORMF} becomes:
\bea\label{FORMFF}
\ds{F_N}&=&\ds{\ap s_{12}\ \lf(\prod_{i=1}^{N-3} \int_0^1 dx_i\ri)\ 
x_{N-3}^{-1+\ap s_{12}}\ (1-x_{N-3})^{\ap s_{23}}}\\[5mm]
&\times&\ds{\prod_{l=1}^{N-4} x_l^{-\ap s_{N-l-1,N-l}+\ap \sum\limits_{j=2}^{N-l-1}s_{1j}+s_{j,j+1}}\ (1-x_l)^{-1+\ap s_{N-1-l,N-l}}\ \ap s_{N-l-1,N-l}\ .}
\eea
As a result of taking the Eikonal limit, the $N-3$ world--sheet integrations
become independent Euler integrals, which integrate to Beta functions
\be\label{FORMFFF}
F_N=\fc{\Gamma\lf(1+\ap s_{12}\ri)\ \Gamma(1+\ap s_{23})}{\Gamma\lf(1+\ap s_{12}+\ap s_{23}\ri)}\ 
\prod_{l=1}^{N-4}\fc{\Gamma(1+\ap x_l)\ \Gamma(1+\ap y_l)}{\Gamma(1+\ap x_l+\ap y_l)}\ ,
\ee
with
\be\label{xyl}\ba{rcl}
\ds{x_l}&=&\ds{s_{N-1-l,N-l}=(k_{N-1-l}+k_{N-l})^2\ ,}\\[5mm]
\ds{y_l}&=&\ds{\sum\limits_{j=2}^{N-l-1}s_{1j}+\sum\limits_{j=2}^{N-l-2}s_{j,j+1}=
\begin{cases}
(k_{N-l}+\ldots+k_N)^2\ ,& l< \floor{\fc{N}{2}}\ ,\\
(k_1+k_2+\ldots+k_{N-1-l})^2\ ,& l\geq  \floor{\fc{N}{2}}\ ,
\end{cases}}
\ea\ee
for $l=1,\ldots,N-4$. E.g. we have
\bea
N=5&:&x_1=s_{34}\ \ ,\ \ y_1=s_{45}\\
N=6&:&x_1=s_{45}\ \ ,\ \ x_2=s_{34}\ \ ,\ \ y_1=s_{56}\ \ ,\ \ y_2=s_{123}\ ,\\
&&\vdots
\eea
with  $s_{ijl}=\ap(k_i+k_j+k_l)^2$.
Note, that the first factor  of \req{FORMFFF} simply represents the four--point result \req{Form4}.
In \cite{Garousi:2010er} for even $N$ a similar expression than \req{FORMFFF} has been considered 
in describing a very restricted and constrained subset of the full kinematics of the SYM factor
of \req{INFOPEN}.

Let us now compute the hard scattering high--energy limit $\ap\ra\infty$  of the result \req{FORMFFF}. In the sequel we apply  the asymptotic formula \cite{Abramowitz}
\be\label{STIRLING}
\Gamma(az+b)\sim (2\pi)^{1/2}\ \exp\lf\{\lf(az+b-\h\ri)\ \ln (az)-az\ri\}\ \ \ ,
\ \ \ |\arg(z)|<\pi,\ a>0\ ,
\ee
to find the following approximation:
\be\label{Apply}
\fc{\Gamma(1+\ap x)\ \Gamma(1+\ap  y)}{\Gamma(1+\ap x+\ap  y)}\sim(2\pi\ap )^{1/2}\ 
\lf(\fc{xy}{x+y}\ri)^{1/2}\ \exp\lf\{\ap \lf[\ x\ln x+ y\ln y- (x+y)\ln(x+y)\ \ri]\ri\}\ .
\ee
With \req{Apply} we now can extract the high--energy limit of the function \req{FORMFFF}
\bea\label{HighF}
\ds{F_N}&\sim&\ds{(2\pi\ap)^{\fc{N-3}{2}}\ \lf(\fc{s_{12}\ s_{23}}{s_{2N}}\ri)^{1/2}\ 
\exp\lf\{\ap (s_{12}\ln s_{12}+s_{23}\ln s_{23}+s_{2N}\ln s_{2N})\ri\}}\\[5mm]
&\times&\ds{\prod_{l=1}^{N-4}\lf(\fc{x_l\;y_l}{z_l}\ri)^{1/2}\ 
\exp\lf\{\ap (x_l\ln x_l+y_l\ln y_l+z_l\ln z_l)\ri\}\ ,}
\eea
with:
\bea\label{zl}
\ds{z_l=-x_l-y_l}&=&\ds{-\sum_{j=2}^{N-l-1}s_{1j}+s_{j,j+1}=\sum_{j=2}^{N-l-1}s_{jN}
+\sum_{j=2}^{N-l-2}s_{j,j+1}}\\[5mm]
&=&\ds{
\begin{cases}
(k_1+k_{N-l}+\ldots+k_{N-1})^2\ ,& l< \floor{\fc{N}{2}}\ ,\\
(k_2+\ldots+k_{N-1-l}+k_N)^2\ ,& l\geq  \floor{\fc{N}{2}}\ ,
\end{cases}}
\eea
for $l=1,\ldots,N-4$.
Note, that in deriving \req{HighF} we have not used the scattering equations \cite{Cachazo}.
We have extracted the limit $\ap\ra\infty$ directly from the explicit expression \req{FORMFFF}.
As a consequence the final result \req{INFOPEN} is given by a single term.

There are two different situations to be discussed. The latter correspond to the two regimes
\req{stringyregime} and \req{fieldregime}, respectively.

\noindent
\underline{Case (i) $\fc{\sqrt s}{N}>M_s:$}

\noindent
For finite $N$ this case is met for small string mass $M_s\ra 0$ (i.e. $\ap\ra\infty$) or large momenta $s\ra\infty$. Then \req{HighF} can be used to approximate the string form factor \req{FORMFFF}. 
With \req{eps} (i.e. finite $\epsilon$) for this region all invariants of the  Eikonal parameterization \req{ER} are of the same order and can be approximated by \req{VAR3}.
For the parameterization \req{VAR3}, i.e. $|s_{ij}|\sim s$
we roughly have
\bea\label{XYZL}
x_l&\sim& s\ \ \ ,\ \ \ y_l\sim-s\ ,\\
z_l&\sim&-s\ \ \ ,\ \ \ l=1,\ldots,N-4\ ,
\eea
and the high--energy behaviour $s\ra\infty$ of the $N$--gluon form factor \req{HighF} behaves as: 
\be\label{highForm}
F_N\sim (\ap s)^{\h(N-3)}\ e^{-(N-3)\ \ap s\ln (\ap s)}\ .
\ee
Together with the YM behaviour  $A_{YM}\sim s^{-\h(N-4)}$ (given in \req{FTbev1}) 
we obtain the following high--energy behaviour $s\ra\infty$ of the open superstring $N$--point amplitude \req{INFOPEN} in the Eikonal constraints \req{VAR1} and \req{VAR2}:
\be\label{HIGHg}
\Ac_N\sim g_{YM}^{N-2}\ \ap^{\h(N-3)}\ s^{1/2}\ e^{-(N-3)\ \ap s\ln (\ap s)}\ .
\ee

\noindent
\underline{Case (ii) $\fc{\sqrt s}{N}<M_s:$}

\noindent
Finally, for small $\eps\ra 0$ (corresponding to $N\ra\infty$) the Eikonal parameterization \req{ER}
describes the Eikonal Regge regime. In this regime some of the quantities \req{xyl} vanish
\be\label{trivial}
x_l\sim 0\ ,
\ee
and the string form factor \req{FORMFFF} becomes trivial:
\be\label{Trivial}
F_N=1\ .
\ee
Hence, in the Eikonal Regge regime, the open superstring amplitude \req{INFOPEN} becomes identical in form to the field--theory amplitude:
\be
\Ac(1,\ldots,N)=g_{YM}^{N-2}\ A_{YM}(1,\ldots,N)\ .
\ee
This fact has been conjectured for the MHV case in \cite{Cheung:2010vn}.
For the latter we recover the SYM result \req{ThreeCases}:
\be\label{HIGHGER}
\Ac_N\sim g_{YM}^{N-2}\  \lf(\fc{s}{(N-2)^2}\ri)^{\h(4-N)}\ f(\phi)\ \times\begin{cases}(N-2)^2\ ,
\quad i,j\in\{1,N\}\ ,\\
							   1\ , \quad i\in\{1,N\}\ ,\ j\notin\{1,N\}\ ,\\
							   (N-2)^{-2}\ , \quad  i,j\notin\{1,N\}\ .
                \end{cases}
\ee
The Eikonal Regge regime corresponds to a limit in which the positions $z_1,z_N$ of string vertex operators $V(z_1),V(z_N)$ are close to each other and generate a pomeron vertex operator \cite{Brower:2006ea}.

Note, that the two results \req{HIGHg} and \req{HIGHGER} represent two different
high--energy limits: while in \req{HIGHg} for finite $N$ the large $s$ (or large $\ap$) limit is taken, in \req{HIGHGER} for infinite $N$ ($\eps=0$) the large $s$ limit is considered.

\subsubsection[High--energy limit of $N$--point graviton amplitude]{High--energy 
limit of $\bm{N}$--point graviton amplitude}

Let us now move on to the closed string $N$--point amplitude describing the scattering of $N$ gravitons. A useful 
way of expressing the latter has recently been presented in the works \cite{Stieberger:2013wea,Stieberger:2014hba} in the  form:
\be\label{GRAVITON}
\Mc(1,\ldots,N)=(-1)^{N-3}\ \kappa^{N-2}\  A_{YM}^t\ S_0\ \sv(\Ac)\ .
\ee
Above $\Ac$ is a $(N-3)!$ dimensional vector encompassing the independent open string subamplitudes
\req{OPEN}. Similarly, $A_{YM}$ denotes an $(N-3)!$ dimensional vector of independent SYM subamplitudes. The map $\sv$ denotes the single--valued map, which roughly speaking projects
an open string amplitude such that it describes a closed string amplitude, cf. Ref. \cite{Stieberger:2014hba} for more details.
Furthermore, $S_0$ is a $(N-3)!\times(N-3)!$ matrix given by $S_0=SK$, with
$S$ being the momentum kernel \req{Kernel} and $K$ accounting for the basis change of SYM subamplitudes
$A_{YM}(1,\rho(2,\ldots,N-2),N,N-1)=K_{\rho}^{\ \sigma}\ A_{YM}(1,\sigma(2,\ldots,N-2),N-1,N)$.

One important observation is the fact, that in the Eikonal limit \req{VAR1} 
only the first element
\be
\sigma_N:=(SK)_{11}
\ee
of the matrix product $SK$ is non--vanishing.  E.g. we have:
\bea
\ds{\sigma_4}&=&\ds{\fc{su}{t}\ \ \ ,\ \ \ N=4\ ,}\\[5mm]
\ds{\sigma_5}&=&\ds{\fc{s_{12}\ s_{23}\ s_{34}\ s_{45}}{(s_{12}+s_{23})\ (s_{34}+s_{45})}\ \ \ ,\ \ \ N=5\ ,}\\[5mm]
\ds{\sigma_6}&=&\ds{\fc{s_{12}\ s_{23}\ s_{34}\ s_{45}\ s_{56}\ s_{123}}{(s_{12}+s_{23})\ (s_{45}+s_{56})\ (s_{34}+s_{123})}\ \ \ ,\ \ \ N=6\ .}\\
&&\vdots
\eea
The general expression for  $\sigma_N$ can be given as
\be\label{general}
\sigma_N=\lf(\fc{s_{12}\ s_{23}}{s_{12}+s_{23}}\ri)\ \prod_{l=1}^{N-4}\lf(\fc{x_l\;y_l}{x_l+y_l}\ri)=(-1)^{N-3}\ \lf(\fc{s_{12}\ s_{23}}{s_{2N}}\ri)\ \prod_{l=1}^{N-4}\lf(\fc{x_l\;y_l}{z_l}\ri)\ ,
\ee
with $x_l,y_l$ given in \req{xyl} and $z_l$ displayed in \req{zl}. 
The function $\sigma_N$ is a rational function
in kinematic invariants of degree $N-3$, i.e. $\sigma_N\sim s^{N-3}$.
Hence, in the Eikonal limit the $N$--graviton amplitude \req{GRAVITON} becomes
\be\label{VeryNice}
\Mc(1,\ldots,N)=\kappa^{N-2}\  |A_{YM}(1,\ldots,N)|^2\ M_N\ ,
\ee
with the form factor
\be\label{withFunction}
M_N=(-1)^{N-3}\ \sigma_N\ \sv(F_N)\ ,
\ee
and  the function $F_N$ given in \req{FORMFFF}.
To extract the high--energy limit $\ap\ra\infty$ of the latter we use \cite{Stieberger:2014hba}
\be\label{Applya}
\sv\lf(\fc{\Gamma(1+x)\ \Gamma(1+y)}{\Gamma(1+x+y)}\ri)=-\fc{\Gamma(x)\ \Gamma(y)\ \Gamma(-x-y)}{\Gamma(-x)\ \Gamma(-y)\ \Gamma(x+y)}\ ,
\ee
and
\be
\fc{\Gamma(\ap x)\ \Gamma(\ap y)\ \Gamma(-\ap x-\ap y)}{\Gamma(-\ap x)\ \Gamma(-\ap y)\ \Gamma(\ap x+\ap y)}\sim 
\exp\lf\{\ap\lf[\ 2x\ln x+2y\ln y-2(x+y)\ln(x+y)\ \ri]\ri\}\ \ \ ,\ \ \ x,y\ra\infty
\ee
to arrive at
\bea\label{highG}
\ds{M_N}&\sim&\ds{(4\pi\ap)^{N-3}\  \lf(\fc{s_{12}\ s_{23}}{s_{2N}}\ri)\ 
\exp\lf\{\fc{\ap}{2}\ 
\lf(\ s_{12}\ln s_{12}+s_{23}\ln s_{23}+s_{2N}\ln s_{2N}\ \ri)\ri\}}\\[5mm]
&\times&\ds{\prod_{l=1}^{N-4}\lf(\fc{x_l\;y_l}{z_l}\ri)\ 
\exp\lf\{\fc{\ap}{2}\ \lf(\ x_l\ln x_l+y_l\ln y_l+z_l\ln z_l\ \ri)\ri\}\ ,}
\eea
with $x_l,y_l$ introduced in \req{xyl} and $z_l$ defined in \req{zl}.

Again, there are two different cases to be discussed. The latter corresponds to the two regimes
\req{stringyregime} and \req{fieldregime}, respectively.

\noindent
\underline{Case (i) $\fc{\sqrt s}{N}>M_s:$}

\noindent
For finite $N$ this case is met for small string mass $M_s\ra 0$ (i.e. $\ap\ra\infty$) or large momenta $s\ra\infty$. Then, \req{highG} can be used to approximate the string form factor \req{withFunction}. 
With \req{eps} (i.e. finite $\epsilon$) for this region all invariants of the  Eikonal parameterization \req{ER} are of the same order and can be approximated by \req{VAR3}.
For the parameterization \req{VAR3}, i.e. $|s_{ij}|\sim s$ and \req{XYZL} 
the high--energy behavior $s\ra\infty$ of the  $N$--graviton form factor \req{highG} behaves as: 
\be\label{HighForm}
M_N\sim (\ap s)^{N-3}\ e^{-\fc{\ap}{2} (N-3)\ s\ln (\ap s)}\ .
\ee
Together with the YM behavior  $A_{YM}\sim s^{-\h(N-4)}$ (given in \req{FTbev1}) 
we obtain the following high--energy behavior $s\ra\infty$ of the closed superstring $N$--point amplitude \req{INFOPEN} in the Eikonal constraints \req{VAR1} and \req{VAR2}:
\be\label{HIGHGg}
\Mc_N\sim \kappa^{N-2}\  \ap^{N-3}\ s\ e^{-\fc{\ap}{2} (N-3)\ s\ln (\ap s)}\ .
\ee
It is interesting to note, that \req{HighForm} essentially is the square of the open string case \req{highForm} subject to a rescaling of the string tension $\ap$ as $\ap\ra\ap/4$. This fact becomes feasible by the single--valued projection \cite{Stieberger:2014hba}.
The limit discussed above corresponds to the stringy region  \req{stringyregime}.
It is important to note that for large $s$ (or small string scale $M_s$) the high--energy  limit of the $N$-point graviton string amplitude is exponentially suppressed in contrast to the
corresponding field theory amplitude. 

\noindent
\underline{Case (ii) $\fc{\sqrt s}{N}<M_s:$}

\noindent
Finally, in the Eikonal Regge regime  $\eps\ra 0$ (corresponding to $N\ra\infty$) some of the quantities \req{xyl} vanish \req{trivial}. In this limit the factor \req{general} scales
as $\sigma_N\sim \lf(\fc{s}{(N-2)^2}\ri)^{N-3}$. As a consequence the whole string form factor \req{withFunction} becomes
\be
M_N= \lf(\fc{s}{(N-2)^2}\ri)^{N-3}\ ,
\ee
and the gravitational string amplitude \req{VeryNice} becomes identical to the corresponding field--theory amplitude \req{BohrKLT}:
\be\label{resi}
\Mc(1,\ldots,N)= \Mc_{FT}(1,\ldots,N)\ .
\ee
Note, that this limit corresponds to the region  \req{fieldregime}.
Hence, in the Eikonal Regge regime the closed superstring amplitude becomes
the field--theory graviton amplitude. 
For the MHV case this fact has also been conjectured in \cite{Cheung:2010vn}.  In the MHV case we recover the explicit 
field--theory expression \req{Nscaling}.

\subsection{Scattering equations and classicalization high--energy limit}
\def\fa{{\frak a}}
\def\fb{{\frak b}}

In this subsection we shall show that, in a parameterization of the classicalization limit, the scattering equations \cite{Cachazo}
 can be solved exactly allowing us to obtain a closed expression for the high--energy limit of the open and  closed superstring tree--level scattering amplitudes for  a generic number $N$ of external legs. In addition, we obtain compact expressions for the field--theory $N$--gluon and $N$--graviton
 amplitudes in the classicalization limit.

\subsubsection{Saddle point approximation and scattering equations}

The generic expression for an open string $N$--point form factor  is given by the real
iterated disk integral (cf. eq. \req{stringBCJ})
\bea\label{Start}
\ds{Z_\pi(\rho)}&:=&\ds{ Z_\pi(1,\rho(2,\ldots,N-2),N,N-1)}\\
&=&\ds{V_{\rm CKG}^{-1}\ 
\int\limits_{D(\pi)} \lf(\prod_{l=1}^N dz_l\ri) \   \fc{\prod\limits_{i<j}^{N} |z_{ij}|^{\ap s_{ij}}}{  z_{1\rho(2)} z_{\rho(2),\rho(3)} \ldots z_{\rho(N-3),\rho(N-2)}z_{\rho(N-2),N}z_{N,N-1}z_{N-1,1}}} 
\eea
specified by some ordering of $N$ points as 
$D(\pi)=\{z_j\in \IR\ |\ z_1<z_{\pi(2)}<\ldots<z_{\pi(N-2)}<z_{N-1}<z_N\}$ (cf. comment below eq. \req{fix}) and the permutations 
$\rho,\pi\in S_{N-3}$.
Furthermore,  the Koba--Nielsen factor $\prod\limits_{i<j}^N|z_i-z_j|^{\ap s_{ij}}$ with the  kinematic invariants \req{kininv} enters in the integrand.
In \req{Start} the factor $V_{\rm CKG}$ 
accounts for the volume of the conformal Killing group of the disk after choosing the conformal gauge. 
It will be canceled by fixing three vertex positions $z_i,z_j,z_k$, i.e. 
$V_{\rm CKG}=\fc{dz_idz_jdz_k}{z_{ij}z_{jk}z_{ki}}$. The
factor $z_{ij}z_{jk}z_{ki}$  can be identified as the standard reparametrization
ghost correlator. 

For fixed--angle scattering, the high--energy limit $\ap\ra\infty$ of the disk integral \req{Start} can be determined  by performing a saddle--point approximation \cite{Fedo}.
Rewriting the Koba--Nielsen factor of the integrand of \req{Start} as 
$$\prod_{i<j}^N|z_{ij}|^{\ap s_{ij}}=\exp\lf\{\fc{\ap}{2}\sum_{i\neq j}s_{ij}\ln|z_{ij}|\ri\}$$ 
yields the saddle point equations
\be\label{SEQ}
\sum_{j\neq i}^N\fc{s_{ij}}{z_i-z_j}=0\ \ \ ,\ \ \ i=1,\ldots,N\ ,
\ee
whose $(N-3)!$ solutions determine the locations
\be\label{SEQS}
\{z^{(l)}_1,\ldots,z^{(l)}_N\}\in {\bf C}\ \ \ ,\ \ \ l=1,\ldots,(N-3)!
\ee 
of the saddle points. Note, that the stationary points \req{SEQS} do not have to lie\footnote{Their actual positions depend on the choice of kinematic invariants \req{kininv}.} within the 
real integration region $D(\pi)$, but may also be complex.
By Cauchy's theorem the saddle point approximation then implies the continuous deformation of the integral 
along $D(\pi)$ (without leaving the domain of analyticity of the integrand) to a new (admissible) 
contour $C_\pi$ (saddle contour) having the same endpoints as $D(\pi)$ and passing through the stationary points \req{SEQS} in the direction of the steepest descent
of $\Re\lf(\sum\limits_{i\neq j}s_{ij}\ln|z_{ij}|\ri)$ \cite{Fedo}. 
Then, the maximum of the integrand is assumed at the isolated points and the full contribution to the asymptotic expansion of the original integral \req{Start} is obtained by adding the amounts 
(of the integrals over small arcs containing these points) from all relevant saddle points \req{SEQS}. Eventually, the saddle points
\req{SEQS} enter the disk integral \req{Start} as
\bea\label{Saddlei}
\ds{Z_\pi(\rho)}&=&\ds{ \lf(\fc{2\pi}{\ap}\ri)^{\fc{N-3}{2}}\ V_{\rm CKG}^{-1} \int_{C_\pi} \lf(\prod_{l=1}^N dz_l\ri) 
(\det{}^{'}\Phi)^{1/2}\ \prod_{a=1}^N{}^{'}\delta\lf(\sum_{b\neq a}^N\fc{s_{ab}}{z_a-z_b}\ri)}\\[6mm]
&\times&\ds{
\fc{\prod\limits_{i<j}^N|z_i-z_j|^{\ap s_{ij}}}{z_{1\rho(2)} z_{\rho(2),\rho(3)} \ldots z_{\rho(N-3),\rho(N-2)}z_{\rho(N-2),N}z_{N,N-1}z_{N-1,1}}
+\Oc(\ap^{-1})\ ,}
\eea
with the Jacobian:
\be\label{detPhi}
\Phi_{ab}=\h\ \fc{\p^2}{\p z_a\p z_b}\sum_{i\neq j}s_{ij}\ln|z_{ij}|=
\begin{cases}
\fc{s_{ab}}{z_{ab}^2}&a\neq b\ ,\\[2mm]
-\sum\limits_{c\neq a}\fc{s_{ac}}{z_{ac}^2}&a=b\ .
\end{cases}
\ee
Of the latter a specific minor $|\Phi|_{pqr}^{ijk}$, arising after deleting
three rows $p,q,r$ and three columns $i,j,k$ of the matrix $\Phi$,  enters 
the determinant $\det{}^{'}\Phi$ as: 
\be
\det{}^{'}\Phi=\fc{|\Phi|_{pqr}^{ijk}}{(z_{ij}z_{jk}z_{ki})\ (z_{pq}z_{qr}z_{pr})}\ .
\ee
Furthermore, in \req{Saddlei} there is the  product of delta--functions
\be
\prod_{a=1}^N{}^{'}\delta\lf(\sum_{b\neq a}^N\fc{s_{ab}}{z_a-z_b}\ri)=z_{ij}z_{jk}z_{ki}\ \prod_{a\neq i,j,k}\delta\lf(\sum_{b\neq a}^N\fc{s_{ab}}{z_a-z_b}\ri)\ ,
\ee
which is independent on the choice $i,j,k$ and hence permutation invariant.
Eventually, \req{Saddlei} can be written as
\bea\label{Saddle}
\ds{Z_\pi(\rho)}&=&\ds{ 
\lf(\fc{2\pi}{\ap}\ri)^{\fc{N-3}{2}}}\\
&\times&\ds{\sum^{(N-3)!}_{l=1}
 [\det{}^{'}\Phi(z^{(l)})]^{-1/2} \fc{\prod\limits_{i<j}^N|z^{(l)}_i-z^{(l)}_j|^{\ap s_{ij}}}{z^{(l)}_{1\rho(2)} z^{(l)}_{\rho(2),\rho(3)} \ldots z^{(l)}_{\rho(N-3),\rho(N-2)}z^{(l)}_{\rho(N-2),N}z^{(l)}_{N,N-1}z^{(l)}_{N-1,1}}+\Oc(\ap^{-1}).}
\eea
The world--sheet string integral \req{Saddlei} is dominated by the contributions of saddle points \req{SEQS} yielding the sum \req{Saddle}. Although the latter may be complex their total contributions to the sum \req{Saddle} must sum up to a real value.

In \cite{Gross:1987kza} the open string saddle points are obtained from saddle points of the closed string scattering by some reflection principle. On the other hand, 
by the single--valued projection \cite{Stieberger:2013wea,Stieberger:2014hba}
the high--energy limit of closed world--sheet sphere integrals can be obtained from the analog
limit of open string integrals \req{Saddle}.

The set of equations \req{SEQ} also appears in the context
of describing Yang--Mills theory by twistor string theory \cite{Witten:2004cp} or recently as  so--called scattering equations relating  the space of kinematic invariants \req{kininv} and   
locations of $N$ punctures on the complex sphere \cite{Cachazo}.
Hence, as already pointed out in \cite{Witten:2004cp,Cachazo} there
seems to be a striking relation between Yang--Mills theory and string
theory  at high energies communicated by the equations \req{SEQ}.

Clearly, for $N=4$ eq. \req{SEQ} boils down to \req{seq}.
In the general case there are $N-3$ (independent) non--linear equations \req{SEQ} to be solved and their solutions \req{SEQS} are difficult to find. Yet for $N=5$ explicit expressions for \req{Saddle} 
can still be evaluated in general and for $N=6$ the explicit solution can be written
in $D=4$ in terms of spinor helicity variables \cite{Weinzierl:2014vwa}.

In the high--energy limit in \req{stringBCJ} each integral $Z_\pi(\rho)$ gives rise to a sum \req{Saddle} over $(N-3)!$ saddle points \req{SEQS}.
A similar sum over the $(N-3)!$  solutions \req{SEQS} of the scattering equations \req{SEQ} 
can be used to specify the SYM factors $A_{YM}(\si)$ in \req{OPEN}  as \cite{Cachazo}
\be\label{CYM}
A_{YM}(1,\ldots,N)=\fc{\int\lf(\prod\limits_{l=1}^Nd\sigma_l\ri)}{\rm Vol\ SL(2,{\bf C})}\ 
\prod_{a=1}^N{}^{'}\delta\lf(\sum_{b\neq a}^N\fc{s_{ab}}{\sigma_{ab}}\ri)\ \fc{E_N(\{k,\xi,\sigma\})}{\sigma_{12}\ldots\sigma_{N1}}\ ,
\ee
with $N$ inhomogeneous coordinates $\sigma_l\in{\bf CP}^1$ ($\sigma_{ab}=\sigma_a-\sigma_b$) and $E_N(\{k,\xi,\sigma\})$ given by some Pfaffian encoding
the external gluon kinematics with  momenta $k_i$ and gluon polarizations $\xi_j$.
As consequence  the high--energy limit of the open superstring amplitude \req{OPEN} becomes a double sum over solutions \req{SEQS} of the scattering equations \req{SEQ}
\begin{align}
\Ac(1,\ldots,N)&=g_{YM}^{N-2}\ \lf(2\pi\ap\ri)^{\fc{N-3}{2}}\\ 
&\times\sum_{a,b=1}^{(N-3)!} \fc{\lf(\prod\limits_{i<j}^{N} 
|z^{(a)}_{ij}|^{\ap s_{ij}}\ri)}{\det^{'}\Phi(z^{(a)})^{1/2}}
\    \fc{E_N(\{k,\xi,\sigma^{(b)}\})}{\det^{'}\Phi(\sigma^{(b)})}\  
{\det}'\Psi(\{z^{(a)}\},\{\sigma^{(b})\})+\Oc(\ap^{-1})\ ,\nonumber
\end{align}
with the generalized Hodges' determinant $\det'\Psi$ encoding the KLT kernel \req{KLTKERN} and specified in \cite{Cachazo:2012da,Stieberger:2013nha}.
Eventually, by applying the KLT orthogonality property \cite{Cachazo:2012da,Cachazo} 
\be\label{ORTHKLT}
\det{}^{'}\Phi(\sigma^{(a)})^{-1/2} \det{}^{'}\Phi(\sigma^{(b)})^{-1/2}\ 
{\det}'\Psi(\{\sigma^{(a)}\},\{\sigma^{(b})\})=\delta^{ab}
\ee
of two solutions $a,b$ of the 
scattering equation \req{SEQ}, one can cast the high--energy limit of \req{OPEN} into a single sum over $(N-3)!$ solutions  \req{SEQS}:
\be\label{finsaddle}
\Ac(1,\ldots,N)=g_{YM}^{N-2}\ \lf(2\pi\ap\ri)^{\fc{N-3}{2}}\  
\sum_{a=1}^{(N-3)!} \fc{\lf(\prod\limits_{i<j}^{N} 
|z^{(a)}_{ij}|^{\ap s_{ij}}\ri)}{\det^{'}\Phi(z^{(a)})^{1/2}}\ E_N(\{k,\xi,z^{(a)}\})+\Oc(\ap^{-1})\ .
\ee

In $D=4$ the sum \req{finsaddle} decomposes into $k$ $R$--charge sectors describing $N^{k-2}$MHV amplitudes (with $k$ negative--helicity states) labelled by $k=2,\ldots,N-2$, with each sector having 
$\left({N-3\atop k-2}\right)$ solutions and 
$\sum\limits_{k=2}^{N-2}\left({N-3\atop k-2}\right)=(N-3)!$ \cite{Cachazo:2013iaa}. 
The latter describes the RSVW residua in super--twistor space \cite{Roiban:2004yf,Witten:2004cp}.


Next, for the high--energy limit of the closed superstring $N$--graviton amplitude we start from the expression \cite{Cachazo}
\begin{align}
\Mc(1,\ldots,N)&=\kappa^{N-2}\ V_{\rm CKG}^{-1}\ \lf(\prod_{j=1}^N\int_{z_j\in \bf C} d^2 z_j\ri)\ \lf(\prod\limits_{i<j}^{N} |z_{ij}|^{\fc{\ap}{2} s_{ij}}\ri)\\
 &\times\sum_{a,b=1}^{(N-3)!} \fc{E_N(\{k,\xi,\sigma^{(a)}\})\ 
E_N(\{k,\tilde\xi,\tilde\sigma^{(b)}\})}{\det^{'}\Phi(\sigma^{(a)})\det^{'}\Phi(\tilde\sigma^{(b)})}
\ {\det}'\Psi(\{z\},\{\sigma^{(a)}\})\ {\det}'\Psi(\{\ov z\},\{\tilde\sigma^{(b)}\}).\nonumber
\end{align}
Note, that the saddle--point method described above, relies on Cauchy's theorem for the integration of 
analytic functions to deform the path of integration $D(\pi)$ 
in the complex plane onto a path of steepest descent. The integration over a domain in 
the multi--dimensional complex plane requires some sort of multi--dimensional generalization 
of the Laplace method \req{Laplace} \cite{Wong}.
For the one--dimensional complex plane ($N{=}4$ case) in eq. \req{highGf}
we have accomplished this by using polar coordinates. 
The saddle points are given by the same equations \req{SEQ} with solutions \req{SEQS}.
After using the KLT orthogonality \req{ORTHKLT} we have \cite{Cachazo}:
\be\label{GRAVSCATT}
\Mc(1,\ldots,N)=\kappa^{N-2}\; \lf(4\pi\ap\ri)^{N-3}\ \sum_{a=1}^{(N-3)!}
 \fc{\lf(\prod\limits_{i<j}^{N} 
|z^{(a)}_{ij}|^{\fc{\ap}{2} s_{ij}}\ri)}{\det^{'}\Phi(z^{(a)})^{1/2}\det^{'}\Phi(\ov z^{(a)})^{1/2}}\ E_N(\{k,\xi,z^{(a)}\})^2+\Oc(\ap^{-1})\ .
\ee

In the previous subsection we have discussed the Eikonal constraints \req{VAR1}. For the latter  
the scattering equations \req{SEQ} separate. 
More precisely, for the region $z_1<\ldots<z_N$, after gauge fixing three positions as \req{fix}
and introducing the parameterization \req{FIX}
the scattering equations \req{SEQ} boil down to the $N-3$ equations 
\be\label{SEQlight}
\ba{rcl}
\ds{\fc{s_{12}}{x_{N-3}}-\fc{s_{23}}{1-x_{N-3}}}&=&0\ ,\\[5mm]
\ds{\fc{-s_{N-l-1,N-l}+\sum\limits_{j=2}^{N-l-1}s_{1j}+s_{j,j+1}}{x_l}-\fc{s_{N-l-1,N-l}}{1-x_l}}
&=&0\ \ \ ,\ \ \ l=2,\ldots,N-2\ ,
\ea\ee
each depending on only one of the remaining $N-3$ positions.
As a consequence in the limit \req{VAR1}  the high--energy behavior $\ap\ra\infty$ of the string form 
factor \req{Start}  is given by a single term in agreement with the results in the previous subsection.

\subsubsection{Solutions of scattering equations in the classicalization high--energy limit}

Interestingly, for special subspaces of kinematics \req{kininv}, the scattering equations \req{SEQ} become Stieltjes sums for zeros of special functions \cite{Stieltjes} and can be solved analytically.
In this subsection we shall see, that a parametrization of the classicalization high--energy limit \req{DvaliRegime} allows for solutions \req{SEQS} of the scattering equations \req{SEQ}
given by the zeros of a Jacobi polynomial.

In units of $\tfrac{s}{(N-2)^2}$ the classicalization high--energy limit \req{DvaliRegime} can qualitatively  be described by the following parametrization 
\bea\label{PARAMETER}
\ds{s_{1,N}}&=&\ds{\h\ (N-3)\ (N-\fa-\fb)\ ,}\\[4mm]
\ds{s_{N-1,N}}&=&\ds{-\h\ (N-3)\ (2-\fb)\ ,\ \ \ s_{1,N-1}=-\h\ (N-3)\ (2-\fa)\ ,}\\[4mm]
\ds{s_{1,i}}&=&\ds{-\h\ (N-2-\fb)\ ,\ \ \ s_{i,N}=-\h\ (N-2-\fa)\ ,}\\[4mm]
\ds{s_{N-1,i}}&=&\ds{\h\ (4-\fa-\fb)\ ,\ \ \ s_{ij}=1\ \ \ ,\ \ \ i,j\in\{2,\ldots,N-2\}\ ,}
\eea
with finite $\fa,\fb$ (e.g. $-1<\fa,\fb<0$). With the identification
\be
\fa=\alpha+N-1\ \ \ ,\ \ \ \fb=\beta+N-1
\ee 
the paramterization \req{PARAMETER} can then be adjusted to:
\bea\label{GPARAMETER}
\ds{s_{1,N}}&=&\ds{\h\ (3-N)\ (\alpha+\beta+N-2)\ ,}\\[4mm]
\ds{s_{N-1,N}}&=&\ds{\h\ (N-3)\ (N-3+\beta)\ ,\ \ \ s_{1,N-1}=\h\ (N-3)\ (N-3+\alpha)\ ,}\\[4mm]
\ds{s_{1,i}}&=&\ds{\h\ (1+\beta)\ ,\ \ \ s_{i,N}=\h\ (1+\alpha)\ ,}\\[4mm]
\ds{s_{N-1,i}}&=&\ds{\h\ (6-2N-\alpha-\beta)\ \ \ ,\ \ \ s_{ij}=1\ ,\ \ \ i,j\in\{2,\ldots,N-2\}\ .}
\eea
For this special two parameter family  of kinematics 
\req{GPARAMETER} (described by $\alpha,\beta$) the scattering equations \req{SEQ} allow for solutions \req{SEQS}, which can be  related to the $N-3$ zeros $x_a,\ a=1,\ldots,N-3$ of the Jacobi polynomial $P_{N-3}^{(\alpha,\beta)}(x)$ \cite{Kalousios:2013eca}. Actually, this solution is degenerate  
by $(N-3)!$, i.e. each solution $z_i^{(l)}=x_{\pi_l(i-1)},\ i=2,\ldots,N-2$ is specified by a permutation $\pi_l\in S_{N-3},\ l=1,\ldots,(N-3)!$ of the $N-3$ zeros $x_a$.
For this solution the SYM amplitude \req{CYM} and the graviton amplitude have been worked out in compact form\footnote{Note, that we have corrected the gauge amplitude by a factor of $\tfrac{1}{(N-3)!}$, which is missing on the r.h.s. of Eq. (11) in \cite{Kalousios:2013eca}.}  \cite{Kalousios:2013eca}
\begin{subequations}
\begin{align}
A_{YM}(1,\ldots,N)&=\sum^{(N-3)!}_{l=1}\fc{1}{\sigma^{(l)}_{12}\ldots\sigma^{(l)}_{N1}}
 \fc{E_N(\{k,\xi,\sigma^{(l)}\})}{\det{}^{'}\Phi(\sigma^{(l)})}\nonumber\\
&=2^{4-\fc{N}{2}}\ (N-3)!!\ \fc{\Gamma\lf(\fc{N-1+\alpha}{2}\ri)\ 
\Gamma\lf(1+\fc{\beta}{2}\ri)\ \Gamma\lf(\fc{N-1+\alpha+\beta}{2}\ri)}{\Gamma\lf(\fc{1+\alpha}{2}\ri)\ 
\Gamma\lf(\fc{N-2+\beta}{2}\ri)\ \Gamma\lf(\fc{2N-5+\alpha+\beta}{2}\ri)}\ H_N(\alpha,\beta)\ ,\label{KaRea}\\[3mm]
\Mc_{FT}(1,\ldots,N)&=\kappa^{N-2}\ \sum^{(N-3)!}_{l=1}
 \fc{E_N(\{k,\xi,\sigma^{(l)}\})^2}{\det{}^{'}\Phi(\sigma^{(l)})}\nonumber\\
&=-\kappa^{N-2}\ 2^{8-N}\ [(N-3)!!]^2\ \fc{\Gamma\lf(\fc{N-1+\alpha}{2}\ri)\ 
\Gamma\lf(1+\fc{\beta}{2}\ri)\ \Gamma\lf(\fc{N-1+\alpha+\beta}{2}\ri)}{\Gamma\lf(\fc{1+\alpha}{2}\ri)\ 
\Gamma\lf(\fc{N-2+\beta}{2}\ri)\ \Gamma\lf(\fc{2N-5+\alpha+\beta}{2}\ri)}\nonumber\\
&\times\fc{\Gamma\lf(1+\fc{\alpha}{2}\ri)\ 
\Gamma\lf(\fc{N-1+\beta}{2}\ri)\ \Gamma\lf(\fc{2N-4+\alpha+\beta}{2}\ri)}
{\Gamma\lf(\fc{1+\beta}{2}\ri)
\Gamma\lf(\fc{N-2+\alpha}{2}\ri)\ \Gamma\lf(\fc{N-2+\alpha+\beta}{2}\ri)}\ H_N(\alpha,\beta)^2\ ,\label{KaReb}
\end{align}
\end{subequations}
respectively. Above, $H_N$ is the helicity dependent part depending on  the external kinematics of momenta $k_i$ and polarizations $\xi_j$ to be specified below.

For $\alpha,\beta>-1$ the $n$--th order Jacobi polynomial $P_n^{(\alpha,\beta)}(x)$ has $n$ distinct
(real) roots in the interval $(-1,1)$. The conditions $\alpha,\beta>-1$ are to be imposed 
 for the orthogonality of the Jacobi polynomials \cite{Szegoe}. However, we may relax these constraints. Therefore, in \req{KaRea} and \req{KaReb} we may consider $\alpha$ and $\beta$ as two distinct arbitrary real parameters:
\be\label{genpara}
\alpha,\beta\in\IR\ .
\ee
In this case $P_n^{(\alpha,\beta)}(x)$ denote generalized Jacobi polynomials \cite{Szegoe}. Note, that  the zeros of the latter, and therefore the solutions of \req{SEQS}, may be complex and the comments below eq. \req{SEQS} apply.
We have verified, that the results \cite{Kalousios:2013eca} can be derived for generic 
parameters $\alpha,\beta\in\IR$ as long as no singularity occurs. So the amplitudes \req{KaRea} and \req{KaReb} are valid for generic parameters $\alpha$ and $\beta$ \req{genpara}.
Hence, we may simply rewrite \req{KaRea} and \req{KaReb} in terms of the parameterisation \req{PARAMETER}
\begin{subequations}
\begin{align}
A_{YM}(1,\ldots,N)&=2^{4-\fc{N}{2}}\ 
\lf(\fc{s}{(N-2)^2}\ri)^{\fc{4-N}{2}}\ (N-3)!!\nonumber\\[1mm]
&\times\fc{\Gamma\lf(\fc{\fa}{2}\ri)\ 
\Gamma\lf(\fc{3}{2}+\fc{\fb-N}{2}\ri)\ \Gamma\lf(\fc{1-N+\fa+\fb}{2}\ri)}{\Gamma\lf(1+\fc{\fa-N}{2}\ri)\ 
\Gamma\lf(\fc{\fb-1}{2}\ri)\ \Gamma\lf(\fc{\fa+\fb-3}{2}\ri)}\ H_N(\fa,\fb)\ ,\label{KAREa} \\[3mm]
\Mc_{FT}(1,\ldots,N)&=-\kappa^{N-2}\  2^{8-N}\ \fc{s}{(N-2)^2}\ [(N-3)!!]^2\ \fc{\Gamma\lf(\fc{\fa}{2}\ri)\ 
\Gamma\lf(\fc{3}{2}+\fc{\fb-N}{2}\ri)\ \Gamma\lf(\fc{1-N+\fa+\fb}{2}\ri)}{\Gamma\lf(1+\fc{\fa-N}{2}\ri)\ 
\Gamma\lf(\fc{\fb-1}{2}\ri)\ \Gamma\lf(\fc{\fa+\fb-3}{2}\ri)}\nonumber\\[1mm] 
&\times\fc{\Gamma\lf(\fc{3}{2}+\fc{\fa-N}{2}\ri)\ 
\Gamma\lf(\fc{\fb}{2}\ri)\ \Gamma\lf(\fc{\fa+\fb-2}{2}\ri)}
{\Gamma\lf(1+\fc{\fb-N}{2}\ri)
\Gamma\lf(\fc{\fa-1}{2}\ri)\ \Gamma\lf(\fc{\fa+\fb-N}{2}\ri)}\ H_N(\fa,\fb)^2\ ,\label{KAREb}
\end{align}
\end{subequations}
respectively.
Above, we have reinstated the $s$--dependence by inspecting \req{ThreeCases} and \req{Nscaling}.
This $s$--behaviour may also be easily extracted from considering the behavior 
of the determinants entering in \req{KAREa} and \req{KAREb}. 
For the $(N-3)\times (N-3)$ reduced matrix  \req{detPhi}
we have $\det{}^{'}\Phi\sim\lf(\tfrac{s}{(N-2)^2}\ri)^{N-3}$, while the determinant of the relevant 
$(N-2)\times (N-2)$ submatrix of $\Psi$ scales as $E_N^2=\det{}^{'}\Psi\sim\lf(\tfrac{s}{(N-2)^2}\ri)^{N-2}$.
Furthermore, we have  the kinematical factor \cite{Kalousios:2013eca}
\bea
H_N(\fa,\fb)&=&\ds{\fc{c_2^{\fc{N}{2}-3}}{\fa+\fb-4}\lf(\fc{2\ (N-3)\ (N-4)\ c_1c_{N-1}c_N}{(2-N+\fa)\ (2-N+\fb)}-c_2c_{N-1}\ \xi_{1,N}\ri)}\\[6mm]
&+&\ds{c_2^{\fc{N}{2}-2}\lf(\fc{c_1\ \xi_{N-1,N}}{2-N+\fb}+\fc{c_N\
    \xi_{1,N-1}}{2-N+\fa}\ri)\ ,}
\eea
with $\xi_{a,b}\equiv \xi_{ab}=\xi_a\xi_b$, $c_1=\xi_{1,i},\ c_{N-1}=\xi_{i,N-1},\ c_N=\xi_{i,N}$ and $c_2:=\xi_{i,j},\ i,j\in\{2,\ldots,N-2\}$. This choice of polarisation vectors (with arbitrary parameters $c_1,c_2,c_{N-1}$ and $c_N$) guarantees
the on--shell condition $\xi_ak_a=0,\ a=1,\ldots,N$ and momentum conservation.

\subsubsection{Fixing combinatorics from scattering equations}

From \req{KAREb} let us now extract the large $N$ behavior (classicalization limit) of the graviton amplitude for
some $-1<\fa,\fb<0$.
First, the kinematical factor $H_N$ behaves as $c^{N/2}$, with some finite constant $c$.
The ratio of Gamma--functions (depending on $N$) can be approximated by \req{STIRLING} as:
$$\fc{\Gamma\lf(\fc{3}{2}+\fc{\fa-N}{2}\ri)\ \Gamma\lf(\fc{3}{2}+\fc{\fb-N}{2}\ri)\ 
\Gamma\lf(\fc{1-N+\fa+\fb}{2}\ri)}{\Gamma\lf(1+\fc{\fa-N}{2}\ri)\ 
\Gamma\lf(1+\fc{\fb-N}{2}\ri)\ \Gamma\lf(\fc{\fa+\fb-N}{2}\ri)}\sim -\lf(\fc{N}{2}\ri)^{3/2}\ .$$
Hence, in total with $[(N-3)!!]^2\sim\sqrt{\tfrac{2}{\pi}}\tfrac{(N-2)!}{\sqrt{(N-2)}}$ we have
\bea\label{ESTI}
\ds{\Mc_{FT}(1,\ldots,N)}&\sim&\ds{\kappa^{N-2}\ 2^{8-N} \ c^N\  \fc{s}{(N-2)^2}\ \lf(\fc{N}{2}\ri)^{3/2}\ [(N-3)!!]^2}\\[3mm]
&\sim&\ds{ \kappa^{N-2}\ \fc{s}{(N-2)^2}\  (N-1)!\ ,}
\eea
in lines with the behavior \req{Nscaling} for the field theory graviton amplitudes.
It is interesting to note, that for $s\sim N$ the number on the r.h.s. of \req{ESTI} approximately coincides with the dimension $(N-3)!$ of the period matrix of the moduli space $\Mc_{0,N}$ of curves of genus zero with $N$ labelled points, which in turn is the set of Riemann spheres with $N$ marked points modulo isomorphisms of Riemann surfaces sending marked points to marked points, i.e. $\Mc_{0,N}\simeq\{(z_1,\ldots,z_N)\in {\bf P}^1({\bf C})\ |\  z_i\neq z_j\}/PSL(2,{\bf C})$.

\subsubsection{High--energy classicalization limit of string amplitudes from scattering equations}
\def\jac{P_{N-3}^{(\alpha,\beta)}(x)}

Stieltjes has already discovered a relation between the zeros of classical polynomials
and the electrostatic equilibrium interpretation of the saddle point approximation, which is closely 
connected with the calculation of the discriminant of these polynomials. 
In fact, in a moment we shall see that in the classicalization parameterization  \req{GPARAMETER} the discriminant of generalized Jacobi polynomials  is related to the Koba--Nielsen factor.
Here, we shall compute the high--energy open  superstring $N$--gluon amplitude \req{finsaddle} 
and the high--energy closed superstring $N$--graviton amplitude \req{GRAVSCATT} in the 
 classicalization parameterization \req{GPARAMETER}. Therefore, we shall 
evaluate \req{finsaddle} and \req{GRAVSCATT} at the solutions of the scattering equations \req{SEQ}, which are described 
by the $N-3$ zeros $x_a,\ a=1,\ldots,N-3$ of the generalized Jacobi polynomial $\jac$.

To proceed we first  need to work out some properties of the zeros $x_a$ of generalized Jacobi polynomials.
With
\be
l=\fc{1}{(N-3)!}\ \fc{\p^{N-3}}{\p x^{N-3}}\ \jac
=\fc{2^{3-N}}{(N-3)!}\ \fc{\Gamma(2N-5+\alpha+\beta)}{\Gamma(N-2+\alpha+\beta)}\ ,
\ee
being the coefficient of the highest term $x^{N-3}$ of the Jacobi polynomial 
$\jac$ the discriminant of the latter is given by \cite{Szegoe}:
\bea
\ds{\Delta_{N-3}}&:=&\ds{l^{2N-8}\ \prod_{1\leq a<b\leq N-3}(x_a-x_b)^2}\\
&=&\ds{2^{-(N-3)(N-4)}\ \prod_{\nu=1}^{N-3}\nu^{\nu-2N+8}\ (\alpha+\nu)^{\nu-1}
\ (\beta+\nu)^{\nu-1}\ (\alpha+\beta+N-3+\nu)^{N-3-\nu}\ .}
\eea
Furthermore, we derive the following identities:
\begin{align*}
\prod_{a=1}^{N-3}(1-x_a)&=(N-3)!\ \fc{P_{N-3}^{(\alpha,\beta)}(1)}{{P_{N-3}^{(\alpha,\beta)}}^{(N-3)}(x)}=2^{N-3}\   \fc{\Gamma(N-2+\alpha)}
{\Gamma(1+\alpha)}\ \fc{\Gamma(N-2+\alpha+\beta)}{\Gamma(2N-5+\alpha+\beta)}\\
&=2^{N-3}\   \prod_{\nu=1}^{N-3}(\alpha+\nu)\ (\alpha+\beta+N-3+\nu)^{-1}\ ,\\
\prod_{a=1}^{N-3}(1+x_a)&=(-1)^{N+1}\ (N-3)!\ \fc{P_{N-3}^{(\alpha,\beta)}(-1)}{{P_{N-3}^{(\alpha,\beta)}}^{(N-3)}(x)}=2^{N-3}\  \fc{\Gamma(N-2+\beta)}
{\Gamma(1+\beta)}\ \fc{\Gamma(N-2+\alpha+\beta)}{\Gamma(2N-5+\alpha+\beta)}\\
&=2^{N-3}\ \prod_{\nu=1}^{N-3}(\beta+\nu)\ (\alpha+\beta+N-3+\nu)^{-1}\ .
\stepcounter{equation}\tag{\theequation}\label{Prelim}
\end{align*}

With these preliminaries for  the kinematic invariants \req{GPARAMETER}, 
the Koba--Nielsen factor of \req{finsaddle} can be worked out for 
any solution \req{SEQS}. The latter is specified by  some permutation 
$\pi_l\in S_{N-3},\ l=1,\ldots,(N-3)!$ acting on  the $N-3$ zeros $x_a$ of the generalized Jacobi polynomials $\jac$ as
$\{z_i^{(l)}=x_{\pi_l(i-1)}\ |\ \ i=2,\ldots,N-2\}$. 
Together with the three $SL(2,{\bf C})$ fixed positions $z_1^{(l)}=-1,\ z^{(l)}_{N-1}=\infty$ and 
$z_N^{(l)}=1$ we obtain 
\begin{align*}
\prod\limits_{i<j}|z_{ij}^{(l)}|^{\ap s_{ij}}&=2^{\ap s_{1N}}\ 
\prod_{a=2}^{N-2}|z^{(l)}_1-z^{(l)}_a|^{\ap s_{1a}}\ 
|z^{(l)}_N-z^{(l)}_a|^{\ap s_{aN}}\ \prod_{2\leq a<b\leq N-2} |z^{(l)}_a-z^{(l)}_b|^{\ap s_{ab}}
\nonumber\\
&=2^{\ap s_{1N}}\ \prod_{a=1}^{N-3}|1+x_a|^{\ap s_{1m}}\ 
|1-x_a|^{\ap s_{mN}}\ \prod_{1\leq a<b\leq N-3} |x_a-x_b|^{\ap s_{mn}}\ ,\nonumber\\
&=\prod_{\nu=1}^{N-3}\ \lf(\fc{\nu^\nu\ (\alpha+\nu)^{\alpha+\nu}
\ (\beta+\nu)^{\beta+\nu}}{(\alpha+\beta+N-3+\nu)^{\alpha+\beta+N-3+\nu}}\ri)^{\ap/2}\ ,
\stepcounter{equation}\tag{\theequation}\label{KOBA}
\end{align*}
with any $m,n\in\{2,\ldots,N-2\}$.
Note, that the above expression is independent on the permutation $\pi_l$ under consideration, i.e.
for the parameterization  \req{GPARAMETER} each solution \req{SEQS} of the scattering equation yields the same Koba--Nielsen factor. In addition, in the sum \req{finsaddle}, the quotient 
$\fc{E_N(\{k,\xi,z^{(a)})}{\det^{'}\Phi(z^{(a)})^{1/2}}$ is independent on the particular
solution $a$. As a consequence we can rewrite this sum  as
\begin{align*}
\sum_{a=1}^{(N-3)!} \fc{\lf(\prod\limits_{i<j}^{N} 
|z^{(a)}_{ij}|^{\ap s_{ij}}\ri)}{\det^{'}\Phi(z^{(a)})^{1/2}}\ E_N(\{k,\xi,z^{(a)}\})
&=(N-3)!\ \lf(|\Phi|^{1,N-1,N}_{1,N-1,N}\ri)^{1/2}\lf(\prod_{a=1}^{N-3}(1+x_a)\ri)  
\lf(\prod\limits_{i<j}|z_{ij}^{(l)}|^{\ap s_{ij}}\ri)\\
&\times\sum^{(N-3)!}_{a=1}\fc{1}{\sigma^{(a)}_{12}\ldots\sigma^{(a)}_{N1}}\ 
 \fc{E_N(\{k,\xi,\sigma^{(a)}\})}{\det{}^{'}\Phi(\sigma^{(a)})}\ ,
\stepcounter{equation}\tag{\theequation}\label{USE}
\end{align*}
with $l$ denoting any solution.
In \req{USE} the last factor yields the SYM amplitude \req{KaRea}.
On the other hand, based on the results in \cite{Kalousios:2013eca} we have:
\begin{align*}
|\Phi|^{1,N-1,N}_{1,N-1,N}&=[(N-3)!]^2\ \fc{l^3}{P^{(\alpha,\beta)}_{N-3}(1)\ P^{(\alpha,\beta)}_{N-3}(-1)}=(-1)^{N+1}\ 2^{9-3N}\ (N-3)!\\
&\times\fc{\Gamma(1+\alpha)}{\Gamma(N-2+\alpha)}
\ \fc{\Gamma(1+\beta)}{\Gamma(N-2+\beta)}\ \lf(\fc{\Gamma(2N-5+\alpha+\beta)}{\Gamma(N-2+\alpha+\beta)}\ri)^3\ .
\stepcounter{equation}\tag{\theequation}\label{DETPhi}
\end{align*}
With \req{Prelim} this gives:
\begin{align*}
\lf(|\Phi|^{1,N-1,N}_{1,N-1,N}\ri)^{1/2}\lf(\prod_{a=1}^{N-3}(1+x_a)\ri) &=
\sqrt{(-1)^{N+1}\ 2^{3-N}(N-3)!}\\
&\times\lf\{\fc{\Gamma(1+\alpha)}{\Gamma(N-2+\alpha)}\ 
\fc{\Gamma(N-2+\beta)}{\Gamma(1+\beta)}\ \fc{\Gamma(2N-5+\alpha+\beta)}{\Gamma(N-2+\alpha+\beta)}\ri\}^{1/2}\ .
\end{align*}
After putting all expressions together we arrive at the final result of \req{finsaddle}
\begin{align*}
\Ac(1,\ldots,N)&=g_{YM}^{N-2}\ \lf(2\pi\ap\ri)^{\fc{N-3}{2}}\ (N-3)!\  
\prod_{\nu=1}^{N-3}\lf(-\fc{\nu\ (\beta+\nu)(\alpha+\beta+N-3+\nu)}{2\ (\alpha+\nu)}\ri)^{1/2}
\stepcounter{equation}\tag{\theequation}\label{SCATTYM}\\
&\times\prod_{\nu=1}^{N-3} 
\lf(\fc{\nu^\nu\ (\alpha+\nu)^{\alpha+\nu}
\ (\beta+\nu)^{\beta+\nu}}{(\alpha+\beta+N-3+\nu)^{\alpha+\beta+N-3+\nu}}\ri)^{\ap/2}\ A_{YM}(1,\ldots,N)+\Oc(\ap^{-1})\ ,
\end{align*}
with the field--theory gluon amplitude given in \req{KaRea}.
Note, that with the parameterization \req{GPARAMETER} for $N{=}4$  the result \req{SCATTYM}
boils down to \req{startYM} with \req{high4}. Furthermore, the analytic structure of the 
result \req{SCATTYM} is very reminiscent of the functional dependence appearing in \req{qualSelberg}.

Next, let us compute the closed superstring $N$--graviton amplitude in the 
high--energy classicalization parameterization \req{PARAMETER}. We start from the expression
\req{GRAVSCATT}.  For our solutions \req{SEQS} 
the determinants $\det^{'}\Phi(z^{(a)})$ and $\det^{'}\Phi(\ov z^{(a)})$ 
are real quantities \req{DETPhi}. The same is true for the  Shapiro--Virasoro factor.
As a consequence the latter can be expressed as a square root 
of the Koba--Nielsen factor \req{KOBA} and the sum in \req{GRAVSCATT}
can be written as
$$\sum_{a=1}^{(N-3)!}
 \fc{\lf(\prod\limits_{i<j}^{N} 
|z^{(a)}_{ij}|^{\fc{\ap}{2} s_{ij}}\ri)}{\det^{'}\Phi(z^{(a)})^{1/2}\det^{'}\Phi(\ov z^{(a)})^{1/2}}\ E_N(\{k,\xi,z^{(a)}\})^2=\lf(\prod\limits_{i<j}^{N} 
|z^{(l)}_{ij}|^{\fc{\ap}{2} s_{ij}}\ri)\sum_{a=1}^{(N-3)!}
\fc{ E_N(\{k,\xi,z^{(a)}\})^2}{\det^{'}\Phi(z^{(a)})}\ ,$$
with $l$ denoting any solution and the last factor being the field--theory graviton amplitude \req{KaReb}. 
Eventually after putting all expressions together we obtain
\begin{align*}
\Mc(1,\ldots,N)&= \lf(4\pi\ap\ri)^{N-3}\ 
\stepcounter{equation}\tag{\theequation}\label{SCATTGR}\\
&\times\prod_{\nu=1}^{N-3} 
\lf(\fc{\nu^\nu\ (\alpha+\nu)^{\alpha+\nu}
\ (\beta+\nu)^{\beta+\nu}}{(\alpha+\beta+N-3+\nu)^{\alpha+\beta+N-3+\nu}}\ri)^{\ap/4}\ \Mc_{FT}(1,\ldots,N) \ 
+\Oc(\ap^{-1})\ ,
\end{align*}
with the field--theory graviton amplitude given in \req{KaReb}.
Again, with the parameterisation \req{GPARAMETER} for $N{=}4$  the result \req{SCATTGR}
yields  \req{highGf}.


The high--energy limits \req{SCATTYM} and \req{SCATTGR} correspond to the Case (i) discussed in the previous subsection, i.e. $M_s\ra 0$ (and $\ap\ra\infty$) for finite $N$ and large momenta 
$s\ra\infty$. If the parameterization \req{GPARAMETER} is taken in units of $s$, i.e. $|s_{ij}|\sim s$
we can easily reinstate the $s$--dependence in \req{SCATTYM} and \req{SCATTGR} and find agreement with the results \req{HIGHg} and \req{HIGHGg}, respectively.

\section{Black hole dominance} 

\subsection{Black hole dominance and a cross--check by semi--classical estimates}
 
   An useful cross-check of large-$N$ scaling of amplitudes is provided by
 applying them to the production of generic classical states composed of much softer gravitons than a black hole of the same mass.  
    It is obvious that such states are in a very weak $\lambda$ domain and thus the semi-classical estimates are expected to be applicable.   We shall then match the perturbative quantum  and  
 non-perturbative semi-classical estimates.

  Such a matching serves us for a double purpose. First, it enables us to obtain an independent input about the scaling of large-$N$ amplitude. It also shows how the suppression of production of non-black hole classical configurations can be understood from 
  $N$-particle perturbative amplitudes.  This understanding gives a valuable information, as it 
  uncovers the corpuscular quantum nature behind the exponential suppression of the production  of classical configurations, described by soft coherent states, in high energy two-particle collision processes.    
  
   As an example, let us estimate the production rate of a {\it classical}  gravitational wave in the 
   above-discussed graviton-graviton scattering. For simplicity, we shall take 
the wave to be monochromatic, of characteristic wavelength $L$ and the amplitude $A_{cl}$.  For such a monochromatic wave, the classical energy per 
wave-length-cubed is  $E\, = \, A^2_{cl} L$. In order to be both in a weak gravity regime as well as in the 
domain of semi-classical approximation, we shall 
demand that the Schwarzschild radius corresponding to this energy is much shorter than the wave-length, 
$R \, = \,EG_N \, \ll \, L$.   Or equivalently, 
\begin{equation}
   A^2_{cl}G_N \, \ll \, 1\,.  
    \label{weakL}
  \end{equation}    
 The leading behavior of the transition probability to such a classical wave can be reliably estimated in the semi-classical approximation, and is given by,  
 \begin{equation}
  P_{2\rightarrow Wave} \, =  \,  e^{-{A^2_{cl}L^2 \over \hbar}}  \times (coupling-dependent~factor) 
  \,, 
     \label{prob2toWave}
  \end{equation}    
  where the quantity in the exponent is the  Euclidean action, $S_{E} \, = \,  A^2_{cl}L^2$. 
  
    In order to make contact between the perturbative matrix element  (\ref{pert}) and the 
 semi-classical one (\ref{prob2toWave}), we have to translate  
the monochromatic wave in the quantum language.  In this language, the wave is a coherent state 
$|N\rangle_{coh}$
of gravitons of momenta $p = \hbar /L$ and  the average occupation number $N\, = \,
{A^2_{cl}L^2 \over \hbar}$, 
\begin{equation}
|N\rangle_{coh} \, \equiv\, e^{-{N\over 2}} \sum_n \, {N^{{n\over 2}} \over \sqrt{n!}} \, |n\rangle \, ,   
\label{coh}
\end{equation}
where $|n\rangle$ are $n$-graviton Fock states of momenta $p = \hbar /L$.   
Notice, that the condition (\ref{weakL}) is simply  $\lambda \ll 1$, signalling that we are 
in a weak-coupling regime in which gravitons can be treated as free and thus the perturbative amplitudes must be fully applicable. 

  By choosing $L$ and $A_{cl}$ appropriately, we can make the parameter $N$ of the coherent state arbitrarily-large 
  for an arbitrary choice of $E$.  In this way, we can create an {\it arbitrarily-classical}  wave of arbitrarily low or high energy.  
     In particular, $E$ can be chosen  to be ultra-Planckian or  
well below the Planck scale,   without affecting the validity of the classical approximation for the 
final monochromatic wave.  
  This fact suggests that for the estimate of the transition probability we should be able to reliably use both semi-classical as well as perturbative quantum amplitudes, and the two must match to the leading order.

The rest of the analysis is straightforward.  We need to estimate the perturbative $S$-matrix element 
$ |\langle2|S|N\rangle_{coh}|^2_{pert}$ using  (\ref{pert}) and match it with  (\ref{prob2toWave}).  
Notice, that since the Fock states that enter in the coherent state  (\ref{coh})  correspond to different occupation numbers of the same fixed momentum (or wavelength) 
gravitons,  for each choice of this wave-length  only one Fock state from this sum 
matches the center of mass energy of the 
initial $2$-graviton states. This is the state $|n \rangle$ with $n = N = \sqrt{s}/p$.  Correspondingly, only the transition to this particular state is possible. That is, 
$\langle 2| S |n\rangle \, = \, \delta_{n,N} \langle 2| S |N\rangle$, where $\langle 2| S |N\rangle$ is given by 
(\ref{pert}).  We thus obtain,  
\begin{equation}
|\langle 2| S |N\rangle_{coh} |^2\, =  |e^{-{N\over 2}} \, \sum_n \, {N^{{n\over 2}} \over \sqrt{n!}}\, \langle 2| S |n\rangle|^2 \, =  \,   e^{-N} \lambda^{N} \, .  
\label{2incoh}
\end{equation}
Matching this expression with (\ref{prob2toWave})  reproduces the exponential suppression of the classical state. 

The factor $\lambda^N$ reveals an extra suppression, due to weak coupling. 
This is expected, since the  transition must be absent in a free theory.   This  extra suppression 
is absent for the case of black hole production, since $\lambda \, = \, 1$, which is one of the reasons 
of black hole dominance.  The other, as explained, is the enhancement by an $e^{N}$ factor 
due to multiplicity of states at the quantum-critical point.

\subsection{Possible subtleties of the perturbative description} 
 
  We would like to stress the possible subtleties of the perturbative framework we are working in  
  and its validity for black hole physics.   
   {\it  A priory}, it is not obvious that signatures of  black hole formation in two-particle scattering can be captured by 
    perturbative amplitudes.  In particular, by tree-level amplitudes that are suppressed by the powers 
    of some weak coupling, such as, the  gravitational or string coupling.  It could happen that no single 
    class of Feynman diagrams describing such weak coupling expansion can be pin-pointed as a source 
    of  black hole formation in two-particle scattering. The answer instead could require either a 
   full  re-summation of infinite number of diagrams, or even  inclusion of contributions of yet  unknown  non-perturbative processes. 
    
     So what makes us think that black hole formation can be captured perturbatively? 
     
      First, an encouragement comes from the fact  that our results allow to create a link between the production of black holes and other classical objects, composed out of softer gravitons than a would-be black hole at a given $\sqrt{s}$.  In other words, we  identify a kinematical regime in which the questions of reliability 
     of black hole production description  is linked, with the reliability of the description  of production of other classical objects, whose quantum composition can be identified beyond any reasonable doubt.  
     
       However, we are going beyond this link  by postulating that there exists a  part of the information 
       that can be extracted from a class of perturbative diagrams within a properly identified kinematical    
   regime.   These are the $2 \rightarrow N$ transition processes. 
   
     What we are suggesting is that in the process of black hole formation, which in general is expected to be a highly non-perturbative phenomenon,  there exists  a well-defined division between the contribution that can be interpreted in the language of perturbative diagrams
    and the fully non-perturbative one.  What is important is that the knowledge of the latter contribution  is crucial for identifying the former one. 
   In other words, without having the non-perturbative input that black holes represent a $N$-graviton bound-state it would be impossible to look for the perturbative counterpart of the process in the form of $2 \rightarrow N$ scattering.   
      
    Thus, we are postulating that it is meaningful to  represent, schematically, the  black hole formation  
 probability as the sum over probabilities, 
 \begin{equation}
\sum_j \, |\langle 2| S |N\rangle_{pert} |^2\  |\langle N |BH\rangle_j |^2  \, , 
\label{split}
\end{equation}
 with each member of the sum representing a product of perturbative and non-perturbative 
 matrix elements.  
Here the sum over $j$ runs over non-perturbative black hole states $|BH\rangle_j$, with their  multiplicity 
scaling as $e^N$.  Of course,  one can say that such a scaling is expected from the 
black hole entropy counting, and one does not need any microscopic theory for postulating it.
  This is certainly true, but solely knowledge of the multiplicity of unknown hypothetical micro-states is useless for  understanding the mechanism of black hole 
 production.  
  
  The new ingredient is contained in the identification of the  projection  $\langle N |BH\rangle_j$  of these states on a $N$-graviton state. It is this identification what enables to conclude that black hole formation process  includes a perturbative part  in the form of the perturbative amplitude of $N$-graviton production.   
 Of course, drawing such a connection is impossible without a microscopic theory and this is where the black hole corpuscular portrait  enters in our analysis.  Since in this picture black hole represents an 
 $N$ soft graviton bound-state at the critical point, it naturally suggest a significant projection on
 an out-state of $N$ free gravitons of wave-lengths equal to the ones of the black hole constituents.    
 
  The subtle point here is not in accepting such an overlap between the $N$ graviton state and a black hole state,  but rather in the perturbative part of the probability, which assumes that we can reliably estimate the 
  $N$-graviton production in perturbation theory.  Viability of the latter assumption has nothing to do with a particular microscopic theory of a black hole and,  as shown above, is generic for perturbative computation of the production rate of arbitrary $N$-particle states in two-particle collision, including the ones not even remotely related to black holes.  This separation of the issues
  is crucial for understanding the framework we are working in.  
  
   In order  to explain why this latter assumption is so subtle, let us consider the 
   two-particle scattering at ultra-Planckian center of mass energy from a fully non-perturbative 
   corpuscular point of view.  In fact, we can very quickly realize that the initial state can be represented 
   as a genuine two-particle state only at {\it infinite} separation.  At finite separation, $L$, the center of mass energy 
  sources a Newtonian gravitational field $\phi(\vec{x})$, which in the corpuscular language itself represents 
  a coherent state of longitudinal gravitons in which the  gravitons of wavelength $L$ have average occupation number $N = EL_P^2$  \cite{Dvali:2011th, Dvali:2011aa, coherent3, coherent4}.
  Schematicaly, we can write this in the following form,   
    \begin{equation}
|Newton\rangle \, = \,  \sum_{n_{k=0}...n_{k=\infty}}  \, \prod_k  \, e^{-{N_k\over 2}}  \, {N_k^{{n_k\over 2}} \over \sqrt{n_k!}}\, |n_{k=0}, ...n_{k=\infty}\rangle \, ,
\label{cohNewton}
\end{equation}
where $|n_{k=0}, ...n_{k=\infty}\rangle$ are the Fock states with definite occupation numbers of 
longitudinal gravitons of wavenumber $k$ and the summation is taken over all possible 
 distributions  of $n_k$-s.  The  function $N_{k}$ represents the data that
 determine the average occupation number of gravitons of wave number $k$ in the given coherent state. 
 The function $N_{|k|}$  is exponentially decaying for $|k| \, \gg \, 1/L$. 
 The dominant contribution to gravitational self-energy, 
   \begin{equation}  
  E_{grav} \, = \, {\hbar \over L_P^2} \, \int d^3\vec{x} \vec{\nabla} \phi  \vec{\nabla} \phi \, \sim \, 
  {\hbar \over L} \,  {s \over M_P^2}   
  \label{selfenergy} 
 \end{equation}
 is coming from  the modes of momenta $k \sim 1/L$, with their number 
 being,  $N \, = \, {s \over M_P^2}  $.  Notice that this number coincides with the number of black hole constituent gravitons.  The only difference is that the gravitons that are present in  the initial state
 have  
 extremely long wave-lengths and their collective coupling $\lambda$ is negligible. 
Correspondingly, neither they contribute significantly to the energy, nor 
are they capable of forming a bound-state.  
  Nevertheless, the message is that an ultra-Planckian initial state for any finite value of $L$ is secretly 
a multi-particle state that on top of the two source particles contains $N$ additional gravitons.  
  As the system evolves in time, decreasing the separation between the initial two source particles, $L$, the multi-particle nature of the initial state becomes more and more apparent.  The peak of the dominant graviton distribution in the coherent state evolves towards the higher momenta.   The non-perturbative $N$-particle physics becomes 
fully important for $\hbar/L$ of order $\sqrt{s} \over N$. At this stage $\lambda$ becomes order one 
signalling that the constituent gravitons are driven into the quantum critical point at which they 
form the bound-state and Bogoliubov modes become gapless.   

  What we are suggesting in our current analysis is that the above fully non-perturbative evolution can 
 be substituted by a perturbative creation of $N$-graviton state and its projection on a black hole state using the non-perturbative input from the microscopic theory.  
  
    The fact that we are able to cross-check the result by normalizing the amplitude to the creation of 
   a generic $N$-particle state,  indicates that the failure of the above program would imply a problem in the description of the production of the $N$-graviton state in perturbation theory, rather than in the
 projection of such state into a black hole quantum state.   It is interesting that  at the level of  the studied kinematic regimes  the perturbative treatment comes up with the adequate physical results. 
 
  \section{Lessons from gravitational multi--particle amplitudes} 
      
   \subsection{Peculiarities of multi-particle amplitudes in gravity} 
   
    One of the outcomes of our analysis is to reveal a special property of multi-particle gravitational amplitudes in contrast to similar amplitudes in {\it non-derivatively} interacting bosonic theories, such as, for example, 
   in a self-interacting scalar theory  $\alpha_{\phi} \, \phi^4$, with a non-derivative coupling $\alpha_{\phi}$.  

It has been known for some time \cite{N!inscalar}  that multi-scalar production amplitudes in such theories exhibit    
(at least at the threshold of producing $N$ on-shell massive scalars of mass $m_{\phi}$ out of some initial few-particle state, the simplest being a single virtual boson of energy $\sqrt{s} = Nm_{\phi}$ )
 a factorial growth, 
 \begin{equation}
  {\mathcal A}_{1 \rightarrow N} \, \sim \, \alpha_{\phi}^{N/2} \, N! \, , 
  \label{Aphi4}
   \end{equation}    
and a corresponding growth of the cross-section, 
\begin{equation}
  \sigma_{1 \rightarrow N} \, \sim \, { 1 \over N!} |{\mathcal A}_{1 \rightarrow N}|^2 \, \sim \, \alpha_{\phi}^{N} \, N! \, ,
  \label{Crossphi4}
   \end{equation}    
 where we have omitted the phase-space and other irrelevant factors. 
In non-derivatively coupled theory the tree-level coupling  $\alpha_{\phi}$ is momentum-independent 
and the factorial growth violates unitarity at sufficiently large $N$.  Perturbation theory breaks down for 
$N \, \gg \, \alpha^{-1}_{\phi}$.  The physical implications of this phenomenon is not fully understood. 
It may signal inapplicability of the perturbative treatment or even an inconsistency of the theory.  Since, 
this question  
is not the focus of our paper we shall not discuss it further, but rather confront the growth of the scalar amplitude with the analogous factorial behavior in gravity and stress the important differences. 

  Notice that the equation    (\ref{Aphi4}) is very similar to  (\ref{pert}) with the difference that 
  $\alpha$ of gravity  is replaced by the scalar self-coupling $\alpha_{\phi}$. However,  the momentum dependence 
  of the gravitational coupling, $\alpha \, = \, L_p^2s/N^2$,  makes a dramatic difference. In particular,  for large 
  $N$ it overpowers
the factorial growth of diagrams.  The resulting amplitude in gravity is exponentially-suppressed as opposed to the  
factorially-exploding counterpart in non-derivative $\phi^4$ theory.  

  Notice, that the perturbative tree-level amplitudes in gravity and in non-derivative scalar theory  have 
  problems in the opposite domains of $N$, with the dramatic difference that in the problematic domain 
  gravity amplitudes are cured by black holes, whereas in $\phi^4$ theory no obvious helper   
  is visible.  
 
 On the other hand, 
  in $\phi^4$ theory scattering for $N \ll \alpha_{\phi}^{-1}$ is unitary, whereas for gravity tree-level unitarity is 
  violated  for small $N$ and large $s$.  However, as discussed above,  in gravity this very domain is excluded by the black hole quantum portrait, due to  collective effects  of graviton Bose-gas. 
 Thus, the black hole physics prevents us from entering there. 
 
    In contrast, the domain 
 $N \, \gg \, \alpha^{-1}$ in gravity is perturbatively-safe,  since in this domain $\alpha^N \sim 
 \lambda^N e^{-N} /N!$,  
 whereas the analogous domain in $\phi^4$ 
 violates unitarity.  In particular, as we have seen,  in gravity this large-$N$  behavior takes care of the exponential suppression in the production of classical configurations composed 
of gravitons softer than the Schwarzschild radius of a $\sqrt{s}$ mass black hole. 
  
  The property of suppression of multi-particle amplitudes in $N \gg \alpha^{-1}$ domain 
  is expected to be shared by other derivatively-coupled theories, which are also considered as candidates 
  for classicalization. For example,  in a theory  $(\partial_{\mu} \phi \partial_{\mu} \phi)^2$  the effective quartic coupling scales as the fourth power of momentum  and 
 the multi-particle production must be suppressed  in the domain  $N \, \gg \alpha^{-1}$.

\subsection{Perturbative insights into non-perturbative black hole production}

  The former discussion on the factorial growth of the cross section for scalar theories of type $\phi^4$ sheds light on how the perturbative amplitudes can foresee the non perturbative existence of black holes. 
The simplest way to identify non perturbative physics within perturbation theory is to look for the limits of applicability of perturbation theory. As previously discussed a key aspect of the approach to quantum gravity based on classicalization lies in replacing ultra--Planckian $2 \rightarrow 2$ strongly coupled processes, violating unitarity already at tree level, by $2\rightarrow N$ weakly coupled processes where the total center of mass energy $\sqrt{s}$ is equi-distributed into the $N$ soft outgoing gravitons. Irrespectively how large is $\sqrt{s}$ the corresponding process at tree level is, for large enough $N$, well defined perturbatively. Indeed, all vertices involved in the process can be made, tuning $N$, arbitrarily small. However, there is a prize that we need to pay when we proceed in this way, namely the growth of the number of tree Feynman diagrams contributing to the $2\rightarrow N$ process. This growth is at the origin of the factorials discussed in the previous subsection.  The interplay between the effective coupling constant and the 
growth of the number of diagrams sets the regime where weakly coupled perturbative analysis at tree level is reliable.

To fix ideas let us consider a generic bosonic theory where amplitudes $2\rightarrow N$ for arbitrary large value of $N$ are not forbidden by any form of the selection rule. To characterize the theory we need to know the number of vertices involved in the process, the number $C(N)$ of tree level diagrams and the value of the effective coupling $g$ entering into each vertex. If we assume $N$ to be large enough and we consider a three point vertex the number of vertices will be order $N$ and the number of trees will grow with $N$ as
(up to exponents unimportant for this discussion),\footnote{To be more precise if we use Cayley's formula we should expect
$C(N) \, \sim \, N! /2^N$ where the factor $2^N$ depends on the specific assumption that the vertex is a three point vertex. Incidentally, note that the factor $2^{-N}$ is consistent with the similar factor appearing in the string result (\ref{ESTI}) presented at the end of section 5.}  
\begin{equation}
C(N) \, \sim \, N! \, .
\end{equation}
consequently the cross section will behave as, 
\begin{equation}
\sigma_{2\rightarrow N} \, \sim N! \, \alpha^{N}
\end{equation}
for $\alpha \equiv g^2$. Note that  for a $\phi^4$ theory we get (\ref{Crossphi4})). 
The leading dependence of $\sigma_{2\rightarrow N}$ on the center of mass momentum is implicitly contained in the effective coupling $\alpha \equiv g^2$. The effective coupling is defining the interaction in the underlying Lagrangian. This interaction term can define a relevant or an irrelevant operator depending on the spin of the bosonic field involved in the process. The $\phi^4$ case corresponds to the marginal case. If it is a relevant operator (as it will be with a  $g\phi^3$ type of theory) then the effective coupling $\alpha$ will depend on the corresponding momentum transfer $\sqrt{t(N)}$ -- which for the classicalization kinematics (where $\sqrt{s}$ is equi-distributed) is of order $\frac{\sqrt{s}}{N}$ -- as $\frac{1}{t(N)}$. However, if the interaction vertex defines an irrelevant operator, as it is the case for the three-point vertex of gravitons, $\alpha$ goes as $t(N)$. In this case we obtain
\begin{equation}
\sigma_{2\rightarrow N} \, \sim \, N!\ \left(\frac{sL_P^2}{N^2}\right)^N
\end{equation}
that is precisely what we have reached for these amplitudes both in the KLT approach (supplemented
by the results from scattering equations) as well as in the string approach. 

Once we have fixed the effective coupling and its dependence both on $s$ and $N$ we can set the limits of perturbation theory. The perturbative approach to multi particle scattering is reliable only if 
\begin{equation}
\frac{\sigma_{2\rightarrow N +1}}{\sigma_{2\rightarrow N}}  \lesssim 1 \, , 
\label{boundnonP}
\end{equation}
which leads to 
\begin{equation}
\alpha \lesssim \frac{1}{N} \, .
\end{equation}
Although the amplitude in absolute terms may not violate unitarity, the turning point 
indicates that some non-perturbative information must be included for the corresponding value of $N$. 
Thus, for gravity the bound   (\ref{boundnonP}) implies.  
\begin{equation}
N \, \gtrsim \, sL_P^2 \, . 
\end{equation}
This is a very interesting result since 
this bound is telling us that $N$ should be larger or equal to the corresponding black hole entropy (equivalently number of constituents) of a black hole with mass equal to the center of mass energy. It is instructive to see how the limits of applicability of perturbation theory are teaching us about the underlying physics. For large value of $s$  and $N$ much smaller than $s$ in Planck units, the amplitude is obviously violating unitarity. At this point you can wonder if increasing the value of $N$ for the same value of $s$ will improve the situation. What you observe is that while you are in the regime with $N$ much smaller than $s$, increasing $N$ is not making the situation better but worst. In other words in this regime the ratio (\ref{boundnonP}) is bigger than one. This perturbative situation changes only when you reach a critical value of $N$ where the ratio (\ref{boundnonP}) reaches one and starts to decrease. The regime where the ratio (\ref{boundnonP}) is bigger than one is precisely the regime corresponding, in the black hole portrait, to the strong coupling regime with $\lambda$ larger than one. Thus, whenever we violate the above  perturbative bound we enter into a regime that requires, in order  to be analyzed, non-perturbative input. Nicely enough this regime precisely agrees with the region $\lambda \geq 1$,  i.e.,  with the region that is cut out using the non perturbative corpuscular information of the black hole portrait. Moreover the turning point happens precisely when $N$ is equal to the black hole entropy. This makes explicit the way perturbation theory anticipates not only the non perturbative black hole formation but also, as already stressed many times, its corpuscular constituency. In other words the perturbative analysis, both in field theory as well as in string theory, sets the limit of applicability of perturbation theory in the classicalization kinematics precisely at the point where the system of outgoing gravitons reaches the dynamical condition defining the critical point of the black hole portrait. Furthermore perturbation theory encodes information about the black hole existence, despite the fact that for corresponding value of $N$ the amplitude is still unitary in the absolute sense.

Finally let us stress the difference with the case where the three point interaction vertex is a relevant operator. In this case the former bound becomes $s\geq N^3$. This means that we don't have problems for arbitrarily large $s$ and small $N$ but instead for large $N$ and small $s$. This is a key difference with the case of irrelevant operators i.e with the case of gravity.

\section{Outlook: Classicalization and black holes in the light of graviton amplitudes}

In previous sections we have collected some results regarding tree level $N$ graviton amplitudes in the Eikonal-Regge kinematical regime. In this summary section we shall complement the discussion, already initiated in the introduction, on the physical meaning of these findings.

In the field theory context we have focused our attention on  two key issues. First of all, we have analyzed how for ultra-Planckian values of $\sqrt{s}$ the amplitude  is smoothed-out once we increase the number $N$ of outgoing gravitons. This kinematic mechanism of unitarization -- which is at the core of the idea of classicalization -- would be nevertheless completely useless if the contribution of this kinematics to the total scattering rate were very much suppressed. So our second task has been to extract from the concrete expressions of the amplitudes this suppression factor. By using the scattering equations in the classicalization limit this has been accomplished for 
graviton amplitudes \req{newGR} in eq. \req{ESTI}. For large $N$ the graviton scattering matrix element in this kinematic regime depends on $s$ and $N$ as $ \sim \, (\frac{sL_P^2}{N^2})^{N} N!$. From this expression we observe that the amplitude starts to be smoothed-out  for $N \, = \, s L_P^2$. We can interpret this value of $N$ as the unitarity threshold for the given value of $s$.  In other words, 
a slower growth of $N$ in  the double-scaling limit ($s, N \rightarrow \infty$)  violates unitarity.  Indeed, parameterizing the scaling 
as $N^{1+\gamma} \, = \,  (s L_P^2)$, the matrix element in large $N$ scales as 
$\sim N^{\gamma N} e^{-N}$, which for $\gamma \, > \, 0$  blows up for sufficiently large $N$. 
However, notice that the final states obtained in unitarity-violating scaling are precisely the ones excluded by non-perturbative corpuscular 
physics, since they correspond 
to the over-critical region of the graviton bound-state,  with $\lambda \, > \, 1$, since 
$\lambda \, = \, sL_P^2/N$.


The key lesson we learn from the expression of the amplitude is how much this concrete kinematical configuration contributes to the total amplitude. Indeed, for this threshold value the suppression factor is $\sim \, e^{-N}$.  The amplitudes with faster growing values of $N$ with $s$ are more suppressed, while the slower-growing ones, that would naively violate unitarity, are excluded by the non-perturbative many-body physics of soft gravitons. 

 As we have discussed, the physics interpretation of the previous result is quite transparent. The value of $N$ at which the amplitude starts to smooth-out is precisely what would be the Bekenstein--Hawking entropy of a black hole of mass equal to the center of mass energy, i.e., $\sqrt{s}$. Moreover,  the suppression factor is precisely what would be the multiplicity of states of such a black hole according to the corpuscular quantum portrait.  
  The crucial information we extract from here is how the amplitude reveals the microscopic structure of the black hole as being composed of the $N$ soft outgoing gravitons. This is precisely, as already stressed in the introduction, what we expect from the $N$-portrait of black holes as composite systems of soft gravitons. Moreover, the kinematical conditions of the outgoing gravitons for $N \, = \, s L_P^2$ are the ones determining the critical point of the graviton Bose-Einstein condensate. It is this criticality what accounts for the entropy needed to compensate the exponential suppression factor in the form of a large multiplicity of gapless Bogoliubov modes.
  
  As already argued the regime with $N$ larger than the threshold value determined by the perturbative amplitude corresponds to $\lambda \, <\, 1$ and although is not violating unitarity is very much suppressed. From the microscopic point of view we understand the large suppression of this multi-particle kinematics as due to the fact that the system defined by the outgoing gravitons is far from the critical point with a well-defined finite gap for the Bogoliubov modes. The regime with small number of outgoing gravitons violates unitarity and corresponds from the microscopic point of view to the strong collective-coupling regime $\lambda\, > \, 1$. This is the regime that the microscopic non-perturbative dynamics is cutting out and this is the key of the unitarization mechanism through the black hole formation.
  
But what have we learned from the string theory amplitudes in this kinematical regime? In this Eikonal-Regge kinematics it is easy to identify when purely stringy effects become relevant. Indeed,  the effective center of mass energy $\sqrt{s_{i,i+1}}$ between two consecutive final state gravitons goes like
$\sqrt{s_{i,i+1}}\, \sim \, \frac{\sqrt{s}}{N}$. Thus, this partial contribution to the total amplitude becomes sensitive to string effects if
\begin{equation}\label{one}
\frac{\sqrt{s}}{N} \geq M_s \, , 
\end{equation}
for $M_s$ being the string mass scale. In such a case each of the $N-3$ vertical graviton propagators
(see Fig. 7) should be effectively Reggeized. In other words, in this multi Regge kinematics and in the regime (\ref{one}), we should effectively dress each propagator with the Regge factor
\begin{equation}
s_{i,i+1}^{(\alpha' s_{i,i+1})} \ , 
\end{equation}
leading to an overall contribution of the order of $e^{-(N-3)\frac{s}{N^2}\ln(\frac{s}{N^2})}$ with $s$ measured in string units. This estimate is to be compared with the result \req{HIGHGg} from the 
string theory computation.

\begin{figure}[H]
\centering
\includegraphics[scale=0.3]{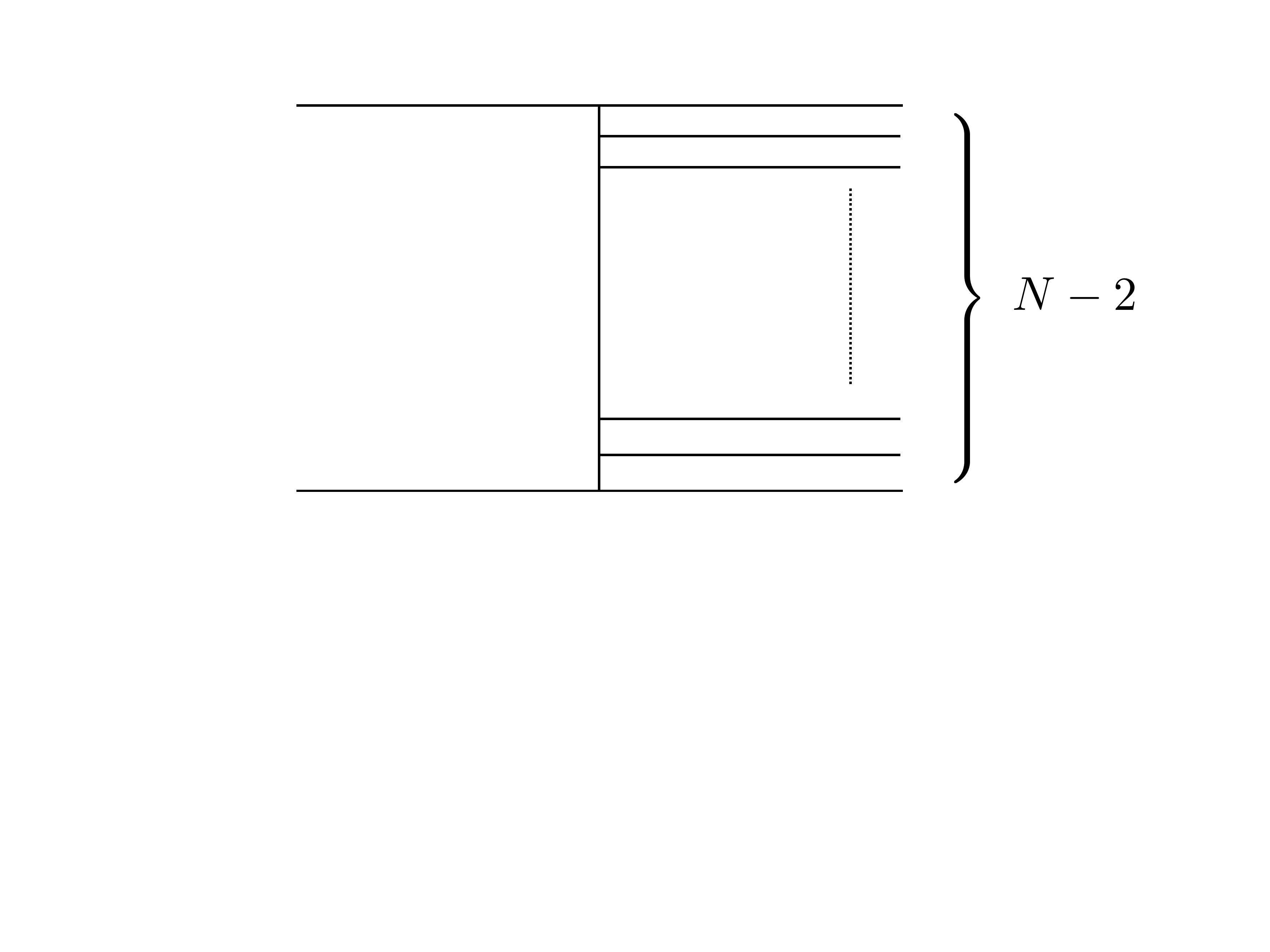}
\vskip-3.5cm
\caption{$N$--graviton scattering  with $N-3$ graviton propagators.}
\end{figure}

Consequently, the field theory computation is reliable if $\frac{\sqrt{s}}{N} \leq M_s$. In this case the factor coming from the Reggeization of the exchanged gravitons becomes one. In section 5 we have considered both situations in the double-scaling limit of both $s$ and $N$ large. In this double-scaling  limit stringy Reggeization effects become relevant for $\frac{\sqrt{s}}{N}$ larger than one (case (i)) while they are suppressed for $\frac{\sqrt{s}}{N}$ smaller than one (case (ii)) (in both cases written in string units). Notice that the concrete value of $L_s$ setting the regime where string effects are relevant only enters, in this kinematics, in the form of the formerly-described Reggeization of the exchanged gravitons.

In order to compare the field theoretic and the string theoretic pictures we need the relation between the two relevant mass scales, namely $L_P$ and $L_s$ given in eq. \req{Planck}.
With this relation  we can, as described already in the introduction, consider different regimes. In the regime where $g_s^2N < 1$ stringy effects due to the Reggeization of the exchanged gravitons starts to be relevant before the created soft gravitons organize themselves into a field theoretic  self-sustained condensate i.e. in the weak coupling regime $\lambda<1$.    For $g_s^2N > 1$ instead the string effects are relevant only in the regime where the outgoing gravitons would be strongly coupled and therefore we could wonder if these Regge effects tame the field theoretical violation of unitarity. 

However, the interesting value at which we want to focus our attention is $g_s^2N =1$, i.e.,  when the threshold of string effects exactly matches the field-theoretic critical point of black hole formation. For this special point we have,
\begin{equation}
g_s \, = \, \frac{1}{\sqrt{N}} \, .
\end{equation}
What is the meaning of this relation? The answer is simply that this value corresponds to the well known string-black hole correspondence. The previous discussion sheds however a new light on this correspondence as determining the point where -- for given kinematics -- the threshold of string effects coincides with the critical point of the graviton Bose-Einstein condensate. Or, as  already stressed in the introduction, this is the situation when at the would-be critical point the string coupling between the constituent quanta becomes equally important as the gravitational coupling \cite{Horowitz:1996nw}     (see also \cite{Dvali:2009ks,Dvali:2010vm}).

Finally,  we would like to put forward a slightly more speculative observation. Until this point although we have been working within the general frame ({\it closed $=$ open $\times$  open}) or equivalently ({\it gravity $=$ YM$^2$ }) we have not used in any explicit way the information about the color of the YM gauge sector. What do we get if we naively use it? As it is customary, we have to use $g_s \, = \, g_{open}^2$, cf. eq. \req{OPCL}. Moreover,  if we think of the open string, as originally pointed out by 't~Hooft, as the planar limit of the gauge theory,  we should identify $g_{open}^2 \, = \, \frac{1}{N_c}$ for $N_c$ being the rank of the gauge group. If we naively combine these two ingredients we arrive to the formal "color-kinematics" relation,
\begin{equation}
N \, = \, N_c^2 \,. 
\end{equation}
Of course in this formal relation $N$ refers to the number of created soft gravitons and thus  it must be interpreted with a bit of care. Note that we arrive to this formal relation only when we put ourselves at the threshold of black hole formation. In these conditions the former relation between $N$ and $N_c$ becomes very reminiscent of the gauge/gravity duality relations. Indeed what this relation does is to identify the black hole entropy $N$ with what would be the $c$--function of the gauge theory. Incidentally, an information that we have never used in our computation of  graviton amplitudes. Pushing a bit forward the analogy, it seems to indicate a deep connection, taking place at the black hole threshold formation, between the hidden Chan-Paton factors dressing the {\it open} string we have used in the computation of the gravitational amplitudes and the gauge holographic dual. Obviously, this observation should be taken with a grain of salt but we feel it certainly deserves a further study.

The former "color kinematics" relation could be anticipated from a different point of view directly working with the gauge theory (or equivalently open string) amplitudes by simply imposing a good planar limit for higher loop amplitudes built using as generalized vertex the gauge theory amplitudes $A_N$. For instance with the gauge theory amplitudes $A_N$ interpreted as effective vertex with $N$ external lines we can define a $2\rightarrow 2$ scattering amplitude with $N-3$ internal loops. This loop gauge amplitude scales with the number of colors as $N_c^{N-3}$ and therefore in order to get a good large $N_c$ planar limit we need to impose (for $N$ large enough)
\begin{equation}
g_{YM}^{2N}s^{-N} N^{2N} \sim N_c^{- N}
\end{equation}
Using now $g^2_{YM}\sim \frac{1}{N_c}$ the former condition becomes $\sqrt{s}=N$ (in open string units) which translated into Planckian language with $g_s= \frac{1}{\sqrt{N}}$ leads to the black hole threshold relation $\sqrt{s} = \sqrt{N}M_P$. In other words what we observe is that at the level of the gauge amplitudes the condition of having a good planar limit in t'Hooft sense (for loop amplitudes) underlies the mechanism of unitarization by black hole formation for the corresponding tree level gravitational amplitudes.

In conclusion, in this paper we have discussed a particular
high--energy limit of Yang-Mills and gravity scattering amplitudes,
both from the field theory as well as from the string
perspective. From the technical side, we have derived new closed
expressions for the tree-level string scattering amplitudes at high energies, which are valid for an arbitrarily large number of external particles. Moreover,  we have considered a particular high--energy limit, which corresponds to the case of classicalization via black hole production, where black holes are bound states of a large number of very soft gravitons. As discussed, this correspondence finds additional support by the existence of a potentially new kind of large $N$ gauge-gravity correspondence with
$N$ the number of external particles in Eikonal Regge kinematics. 
As pointed out it would be interesting to relate this large $N$ duality to the standard large $N$ duality arising in the context of holography and
the AdS/CFT correspondence.

\vskip0.5cm
\goodbreak
\centerline{\noindent{\bf Acknowledgments} }\vskip 1mm
We wish to thank Ram Brustein, Lance Dixon, Martin Sprenger,  and especially Song He for valuable discussions. We thank Wolfgang M\"uck for useful comments and for pointing out some typos in the text. 
This work was partially supported by the 
Humboldt Foundation (under Alexander von Humboldt Professorship of G.D.), ERC Advanced Grant ``Strings and Gravity" (Grant No. 32004), the ERC Advanced Grant ``UV-completion through Bose-Einstein Condensation (Grant No. 339169) and by the DFG cluster of excellence ``Origin and Structure of the Universe", NSF grant  PHY-1316452, FPA 2009-07908, CPAN (CSD2007-00042) and HEPHACOSP-ESP00346, 
TRR 33 The Dark Universe.

\end{document}